\documentclass[prb,twocolumn,showpacs,preprintnumbers,amsmath,amssymb,superscriptaddress]{revtex4}

\usepackage{dcolumn}
\usepackage{bm}
\usepackage{amsmath,graphicx}
\usepackage{color}

\newcommand{\pa}{\partial}

\begin{document}

\title{Morphometric analysis of polygonal cracking patterns in desiccated starch slurries}

\author{Yuri Akiba}
\affiliation{Department of Environmental Sciences, University of Yamanashi, 4-4-37, Takeda, Kofu, Yamanashi 400-8510, Japan}
\author{Jun Magome}
\affiliation{Department of Environmental Sciences, University of Yamanashi, 4-4-37, Takeda, Kofu, Yamanashi 400-8510, Japan}
\affiliation{Interdisciplinary Research Center for River Basin Environment (ICRE),
4-3-11, Takeda, Kofu, Yamanashi 400-8511, Japan}
\author{Hiroshi Kobayashi}
\affiliation{Department of Environmental Sciences, University of Yamanashi, 4-4-37, Takeda, Kofu, Yamanashi 400-8510, Japan}
\author{Hiroyuki Shima}
\email{hshima@yamanashi.ac.jp}
\thanks{(Correspondence author)} 
\affiliation{Department of Environmental Sciences, University of Yamanashi, 4-4-37, Takeda, Kofu, Yamanashi 400-8510, Japan}

\date{\today}

%
%

\begin{abstract}
We investigate the geometry of two-dimensional polygonal cracking that forms on the air-exposed surface of dried starch slurries. Two different kinds of starches, made from potato and corn, exhibited distinguished crack evolution, and there were contrasting effects of slurry thickness on the probability distribution of the polygonal cell area. The experimental findings are believed to result from the difference in the shape and size of starch grains, which strongly influence the capillary transport of water and tensile stress field that drives the polygonal cracking.
\end{abstract}


\maketitle

\section{Introduction}

Desiccation cracks are ubiquitous and exist commonly in everyday life.
Examples include those that develop in dried mud \cite{Kindle1917,Goehring2010,JHLi2011}, 
old paintings \cite{Bucklow1997,Krzemien2016}, ceramic glaze \cite{BohnPRE_Temporal_2005,BohnPRE_From_2005}, and so on.
These cracks usually exhibit a specific network structure, 
splitting the entire surface of the fractured media into many polygonal cells.
Biological analog is found 
in crocodiles' head scales \cite{Milinkovitch2013,ZhaoQin2014},
in which polygonal patterns are produced by a differential growth of tissue;
the leaf venation may further be 
included in the category \cite{Couder2002,Laguna2008,BarSinai2016}.
Despite the variety of chemical compositions and length scales, the apparent similarity in their topology implies the existence of a common governing mechanism that is responsible for the polygonal pattern formation.

Earlier studies have elucidated the mechanism
of desiccation crack formation for various grain-liquid mixtures
\cite{Groisman1994,Colina2000EPJE,Colina2000Mater,Shorlin2000,Lecocq2002,Toramaru2004,Nakahara2006,Bisschop2008,CSTang2011,SNag2010,Nakayama2013,Costa2013,DeCarlo2014,Khatum2015,Nandakishore2016,Kitsunezaki2016}.
The mixtures typically consist of a skeleton of solid grains and pore spaces that are
filled with either liquid or air bubbles.
Upon drying, liquid content evaporates from an air-exposed surface,
which causes shrinkage in the volume of the solidified mixture.
As a consequence, cracks occur when the tensile stress induced by shrinkage 
exceeds the bonding strength of grains.
Recent advances in computer simulation techniques
and theoretical development have enabled us to reproduce real cracking patterns
and to obtain insight into the detailed mechanism.
\cite{Kitsunezaki2010,Maurini2013,Ito2014,Hirobe2016}.

Of the many candidates, corn starch slurry 
({\it i.e.}, a dense mixture of corn starch powder with distilled water)
\cite{Muller1998,Mizuguchi2005,Goehring2005,GoehringPNAS2009,GoehringPRE2009,Crostack2012}
can exhibit heterogeneous crack network characteristics
consisting of two distinct types, namely primary cracks and secondary cracks.
Primary cracks are long and nearly straight cracks,
and they emerge in advance of the appearance of secondary cracks.
Primary cracks originate from the mismatch in the ratio of the 
homogeneous volume shrinkage between the dried slurry and substrate,
as well as the adhesion of the bottom surface of the slurry to the substrate.
In contrast, secondary cracks are relatively short and sinuous cracks, and are
made by spatial inhomogeneity in the local shrinkage of the slurry.
Both types of cracks initially appear at the air-exposed surface,
and they then penetrate the bulk as evaporation takes place in order to
release the internal tensile stress.
In particular, full-depth propagation of the second cracks
results in a regular array of tiny polygonal prisms,
as a reminiscent of columnar joints, a type of impressing geological structure 
that spontaneously occurs in cooling lava flows \cite{Degraff1989,Goehring2008}.

The present work focuses on the geometry in the planar random tessellation 
made by secondary cracks at the surface of desiccated starch slurries.
We note the possibility that the geometry will be affected by
the degree of the irregularities with respect to the size and shape
of the constituent starch grains.
This is because the irregularities cause wide structural variations 
in the pore spaces between adjacent grains;
the pores regulate the local transport of water content,
and play the roles of a channel and a storage pool, depending on the configuration
of grains surrounding 
the pores \cite{GoehringPRE2009,Shokri2010}.
Therefore, from a statistical perspective, it is anticipated that the irregularities 
are responsible for 
local volume shrinkage, and thus for the geometric properties
of the polygonal cracks.
To examine the conjecture,
we performed the desiccation-cracking experiments
using starches made from two different ingredients, potato and corn,
which showed different grain shapes and different-sized 
distributions \cite{Singh2003}.
We conducted image analyses and geometric characterization of the resulting polygonal cracking
in order to evaluate the effect of the size and shape of grains on the crack geometry, as well as to better understand
the underlying mechanism.
We also considered the effect of varying the sample thickness of the starch slurries
on the geometric properties.

\begin{figure}[ttt]
\fbox{\includegraphics[width=3.92cm]{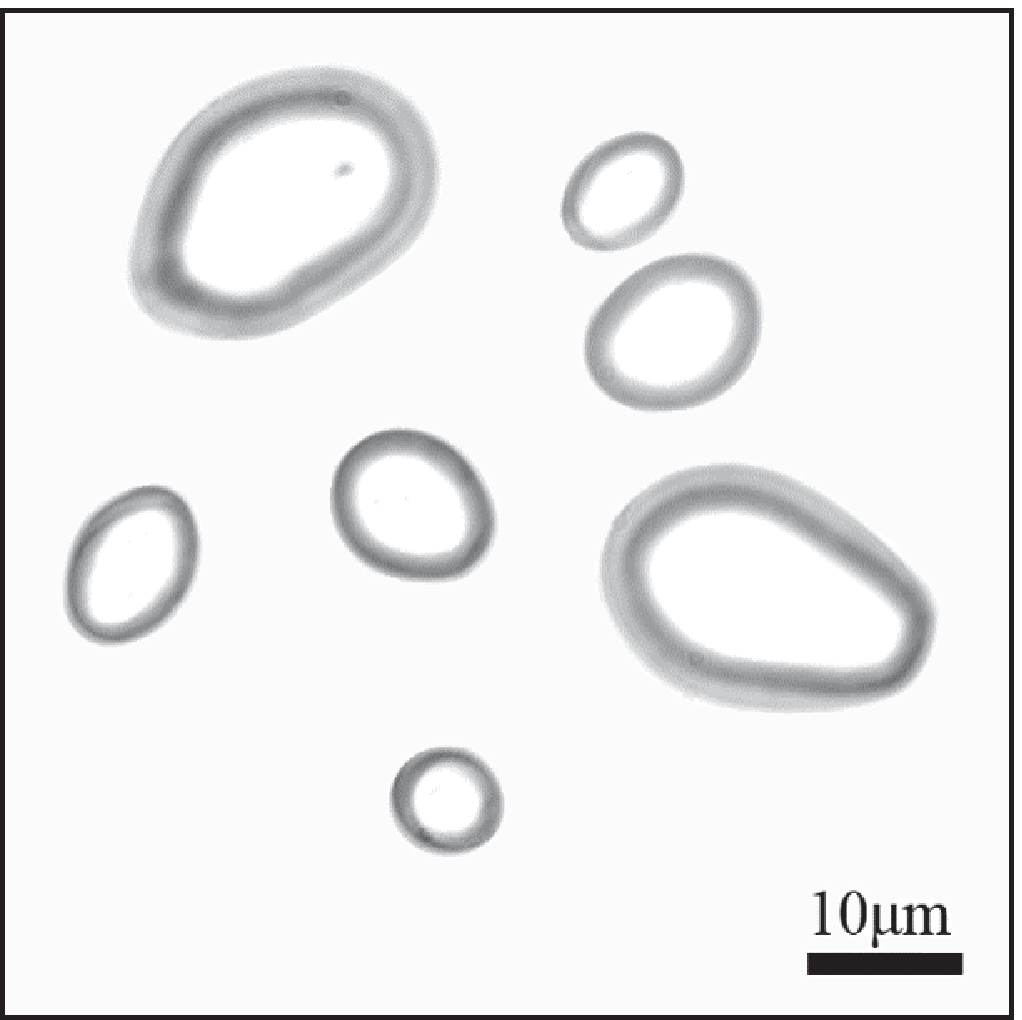}}
\hfill
\fbox{\includegraphics[width=4.03cm]{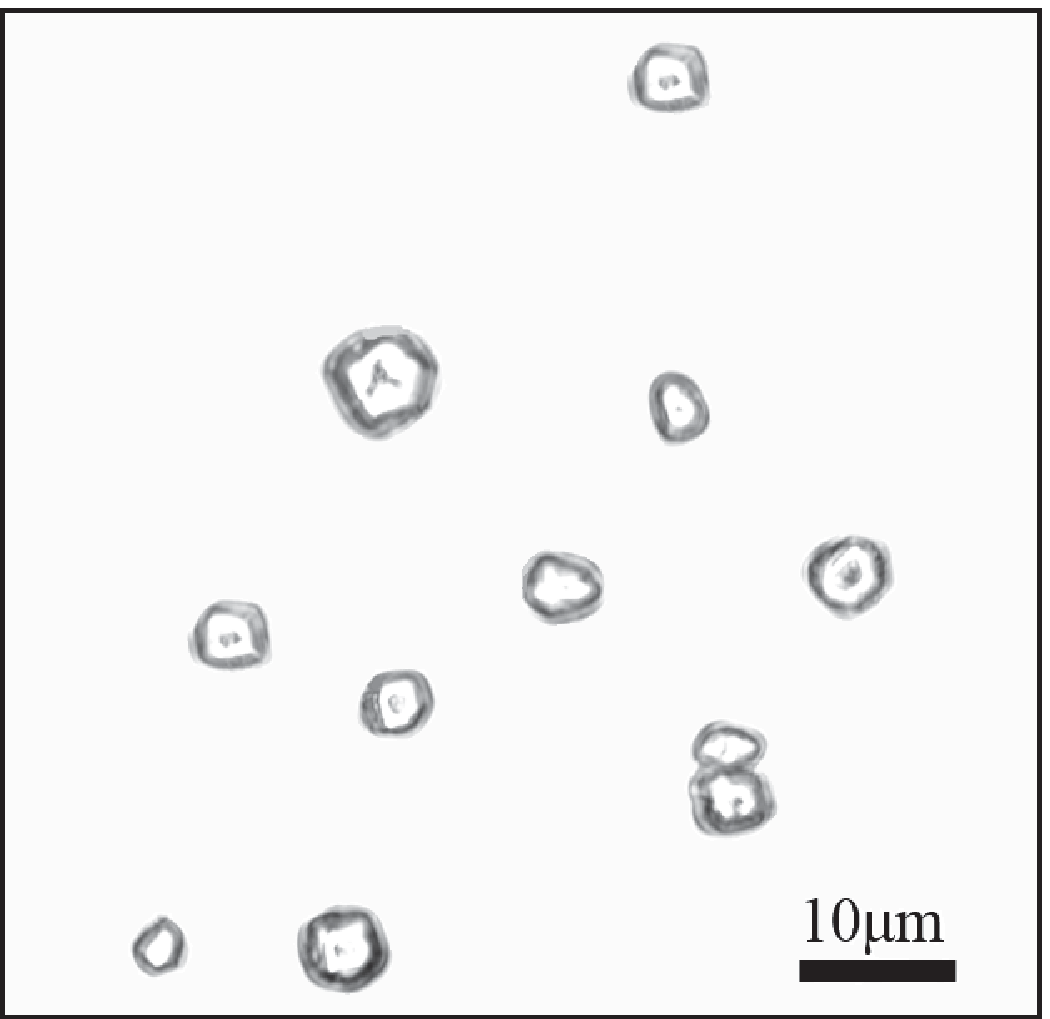}}
\caption{
Optical microscope image of starch grains.
(a) Potato starch. (b) Corn starch.
}
\label{fig:photo}
\end{figure}

\section{Materials and Methods}

\subsection{Slurry sample preparation}

In this study, we used two kinds of commercially available starch powders
made from cone and potato. 
Potato starch is known to exhibit a larger grain size and lower sphericity
compared with those of corn starch,
and we confirmed this by performing optical microscopy measurements
(we will revisit this issue later).
Our sample preparation method was essentially similar to those 
reported in Refs.~[\onlinecite{Muller1998,Toramaru2004,Goehring2005}],
as explained below.

At the initial stage,
saturated slurry samples were prepared by mixing the starch ($w_s$ g in weight) 
with distilled water ($w_d$ g)
at a gravimetric ratio ($=w_s/w_d$) of 55.6\% and 58.3\%
for corn and potato starches, respectively.
The obtained mixture was stirred by hand, 
and the desired quantity was then poured 
into a horizontal container with a circular shape, 
which was 85 mm in diameter and 12 mm in depth. 
In the container, starch grains settled out to the bottom and 
water segregated on the top,
resulting in a dense starch phase covered by a water layer. 
After completion of the settling, the dense starch layer showed thicknesses of either 2 mm, 5 mm, 8 mm, or 11 mm depending on the quantity of the mixture initially poured into the container,
while the thickness of the supernatant clear water was nearly equal or less than 1 mm for all samples.
The decision on the completion of sedimentation was 
based on visual examination;
though a very small amount of starch grains may be contained in 
the apparent transparent water, it was confirmed that subsequent precipitation
of the residual infinitesimal impurities did not give 
a feasible contribution to the sedimentation layer thickness.

We then left the samples untouched until the supernatant water dried out naturally.
Once all of the supernatant water disappeared and 
the top surface of the settled slurry was exposed to air,
the container was placed under a 36-W lamp
and the evaporation time measurement began.
The lamp enabled us to secure reproducible drying condition
as well as to keep the temperature of the exposed surface 
at c.a. 27$^\circ$C against vaporization cooling.
The distance between the top surface of the slurry
and the bottom of the lamp bulb was fixed at 7.0 cm.
All of the experiments were conducted at room temperature; 
26 $\pm$ 2 $^\circ$C, 50 $\pm$ 5\% of moisture humidity.

\subsection{Image analysis technique}

During the evaporation, we took pictures of the drying surface
using a digital camera over a period of time.
We also measured the weight of the starch slurry 
over the same period,
and in this way, we were able to evaluate the amount of water contained in the slurry
and its time variation.
We analyzed the pictures of the polygonal cracks after drying
using image-processing software, called ArcGIS \cite{arcgis},
which is a geographic information system tool employed to analyze geographic data. 
In the image analysis, each image was enhanced in contrast, and was then converted to 
a black-and-white representation. 
We then erased the image of tiny cracks that were isolated from the main network
of the polygonal tessellation as well as those that branched off the network, but which were terminated without connecting to any other cracks.
From the results, we obtained a relatively accurate description of the planar random tessellation
that formed on the top surface of the desiccated starch slurries.
See Appendix A for the details of image processing based on ArcGIS software.

\begin{figure}[ttt]
\includegraphics[width=8.0cm]{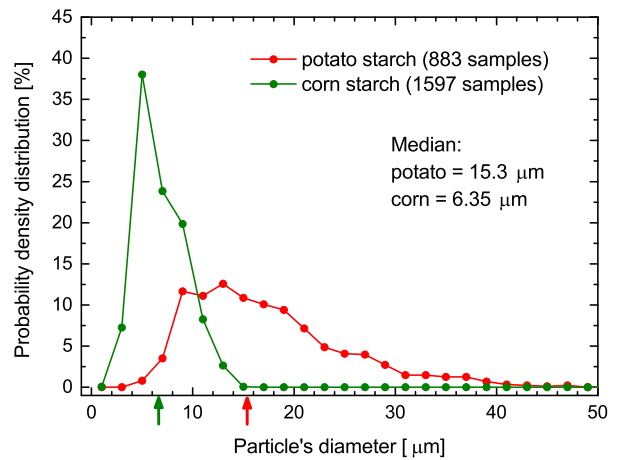}
\caption{
Probability distribution of 
equivalent circular diameter
of starch grains.
}
\label{fig:particlesize}
\end{figure}
\begin{table}[ttt]
  \caption{Elongation factor ($F_{\rm el}$) of starch grains and 
the number of grain samples for averaging.
See Eq.~(\ref{eq_001}) for the definition of $F_{\rm el}$.}
  \label{table:elongationfactor}
  \centering
  \begin{tabular}{ccc}
    \hline
    & Potato starch & Corn starch  \\
    \hline 
    $F_{\rm el}$  & 1.67$\pm$0.15 & 1.54$\pm$0.12 \\
    Num. of samples  & 19 & 21 \\
    \hline
  \end{tabular}
\end{table}

\subsection{How to distinguish two types of cracks}

The two types of cracks can be differentiated in a visual way \cite{Mizuguchi2005,Goehring2010,Crostack2012}. Primary cracks appear successively and they typically meet at T-shaped junctions, since a later crack will preferentially meet an earlier one in a normal direction. They show rectilinear shape with clear, well-defined edges, and quickly penetrate both in the lateral and vertical directions. In contrast, secondary cracks emerge simultaneously, initiated at both the edge of an existing primary crack and a tiny flaw on the air-exposed surface. They propagate slowly with curvy shape, forming Y-shaped or V-shaped branches on the surface. In the vertical direction, they show prismatic structures with polygonal cross sections.

\subsection{Microscopic observation of starch grains}

Prior to performing the desiccation crack experiments,
we examined the microscopic size and shapes of the starch grains
to be used.
Figures \ref{fig:photo}(a) and \ref{fig:photo}(b)
display the optical microscope photographs of starch grains made from
potato and corn, respectively, 
which were obtained using an inverted microscope (CKX31, Olympus)
incorporated with a cooling CCD camera (BU-52LN, Bitran).
The samples taken by photos were obtained by sufficiently 
diluting the starch powder with distilled water.
It is clear from Fig.~\ref{fig:photo}(a) that
the grains of potato starch have smooth round and ellipsoidal shapes with 
a relatively large variation in their sizes.
Meanwhile, the grains of corn starch exhibit slightly angular but nearly spherical shapes, along
with some uniformity in their sizes, as seen in Fig.~\ref{fig:photo}(b).
These morphological features of starch grains are consistent with
earlier observations by scanning electron microscopy (SEM)
\cite{Singh2003,GoehringPRE2009}.
In particular, a high-resolution SEM image of 
corn starch grains unveiled
the less-smooth and highly porous surfaces, which may account for 
the strong hydrophilic qualities of corn starch \cite{GoehringPRE2009}.

The degree of the grain shape's deviation from an ideal sphere can be quantified by
the elongation factor $F_{\rm el}$, which is defined by
\begin{equation}
F_{\rm el} = \frac{ {\rm max}(D_{\rm Fer})}{\ell_{\rm rec}}.
\label{eq_001}
\end{equation}
Here, ${\rm max}(D_{\rm Fer})$ is called 
the maximum 
Feret's diameter \cite{MerkusBook}, which
represents the length of the line segment connecting the two points having the greatest separation
on the perimeter of the particle's cross-section.
$\ell_{\rm rec}$ is the length of the short side of the rectangle 
that has the same perimeter and area as those of the particle's cross-section.
In other words,
the more elongated the shape of a particle, the higher is its elongation factor $F_{\rm el}$.
Table \ref{table:elongationfactor} presents 
the values of $F_{\rm el}$ for potato and corn starch that
average over c.a. 20 grain samples.
Potato starch showed a larger value of $F_{\rm el}$ than corn starch,
and this is attributed to the more elongated grain geometry.

\begin{figure*}[ttt]
\includegraphics[width=4.2cm]{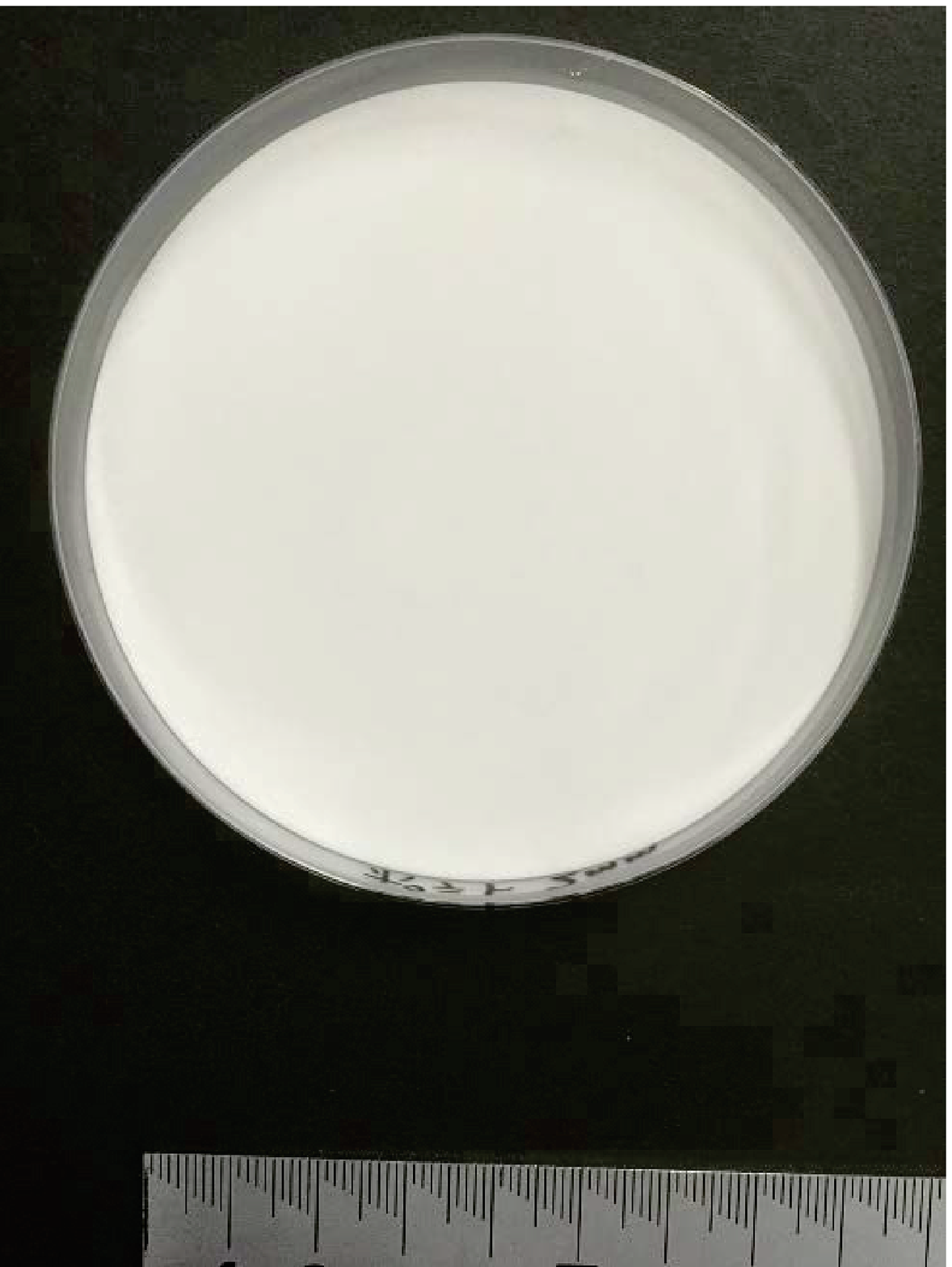}
\includegraphics[width=4.2cm]{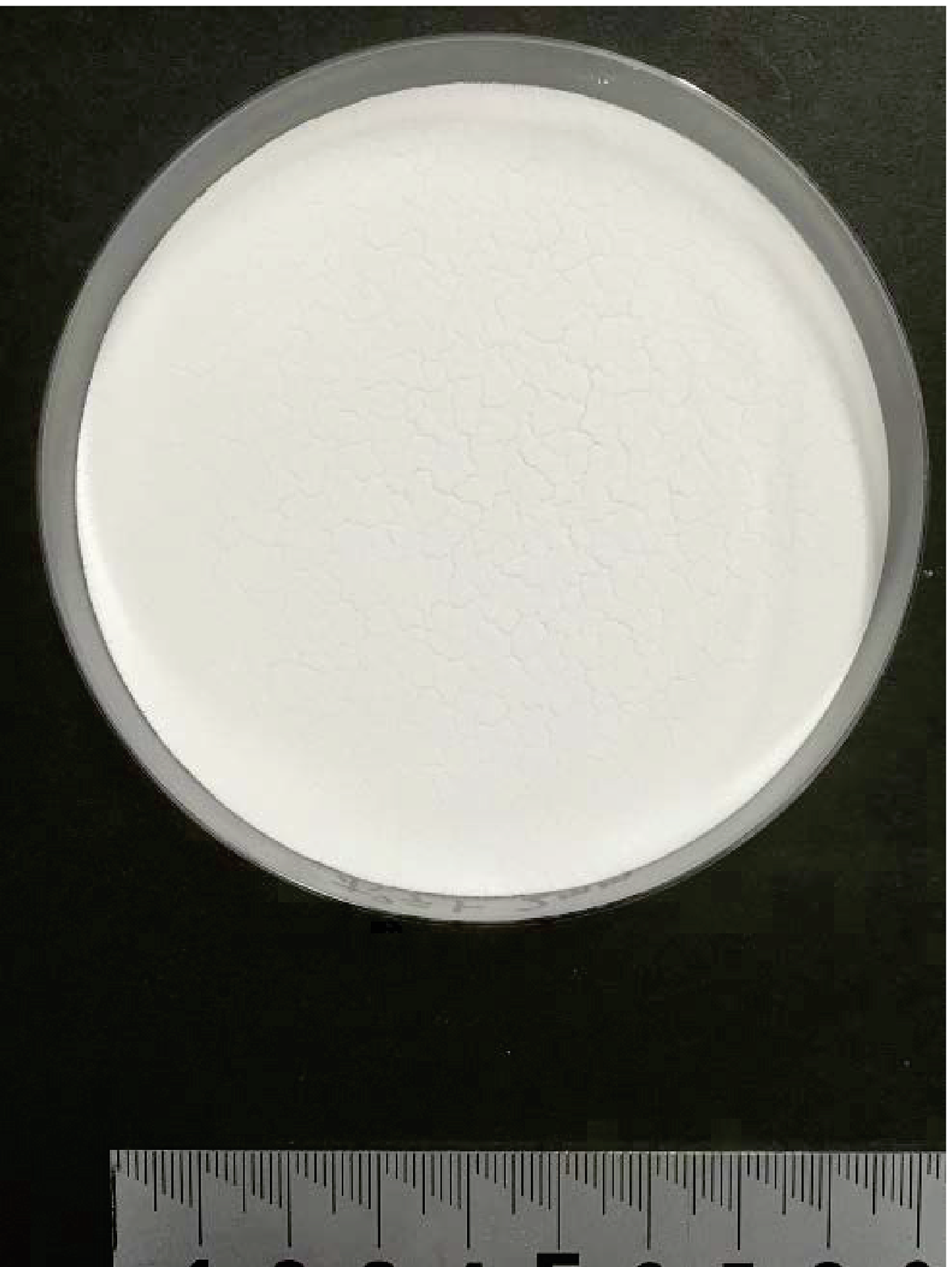}
\includegraphics[width=4.2cm]{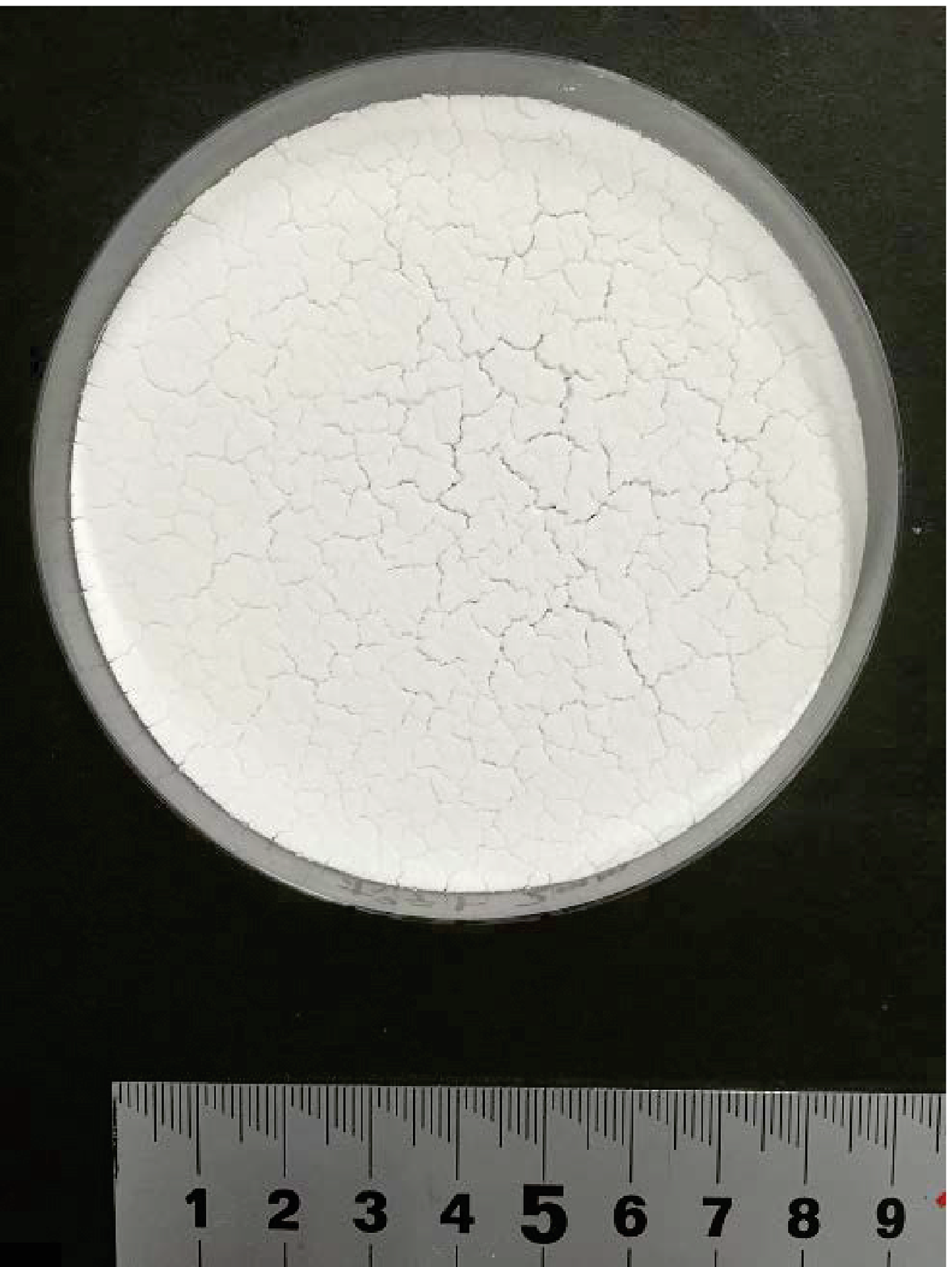}
\includegraphics[width=4.2cm]{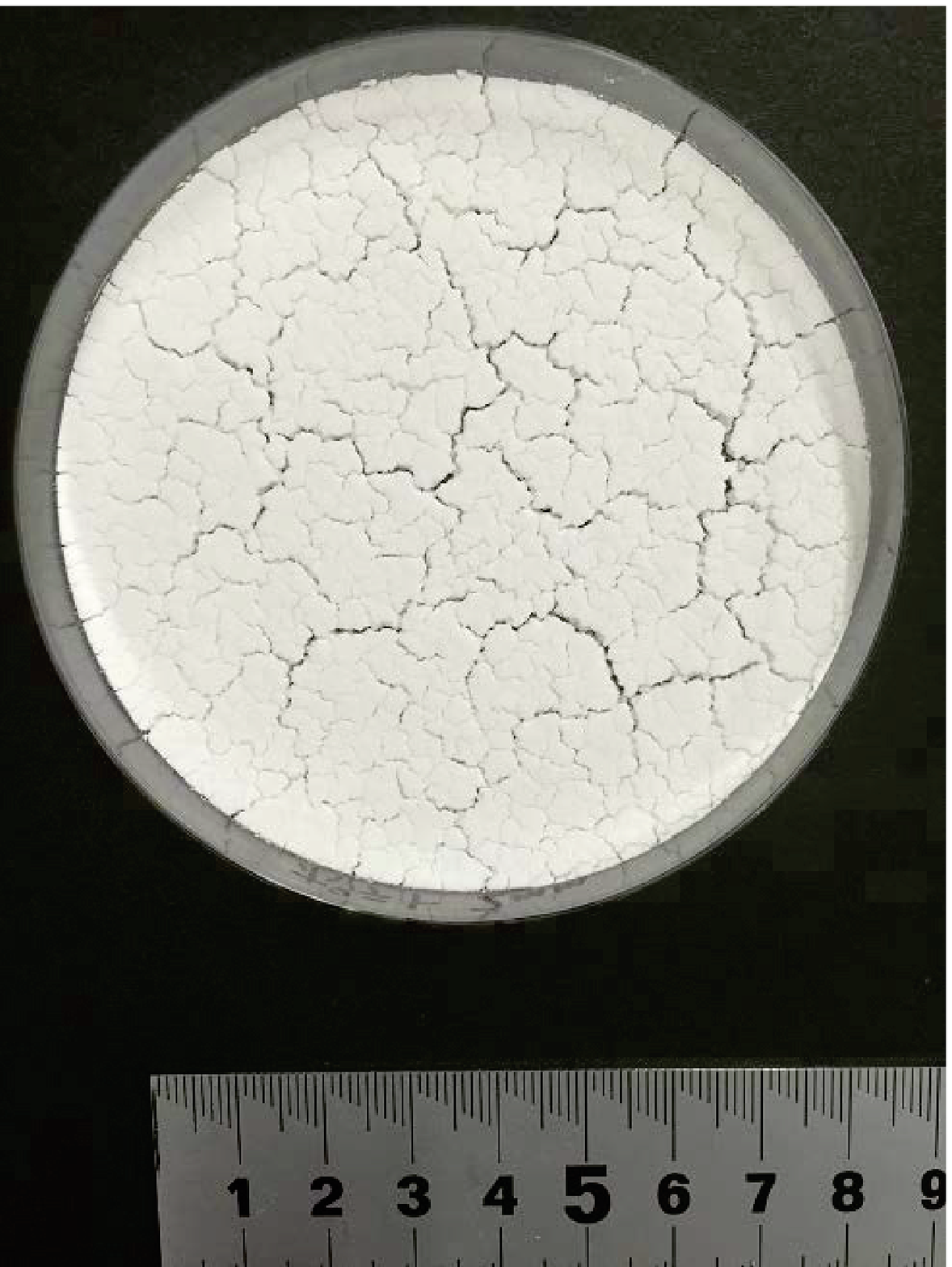}
\caption{
Potato starch slurries with a 2-mm-thick layer 
showing secondary crack propagation over the dried surface.
The time of drying is 60 min, 150 min, 270 min, and 360 min from left to right. 
}
\label{fig:potatocrackphoto}
\end{figure*}
\begin{figure*}[ttt]
\includegraphics[width=4.1cm]{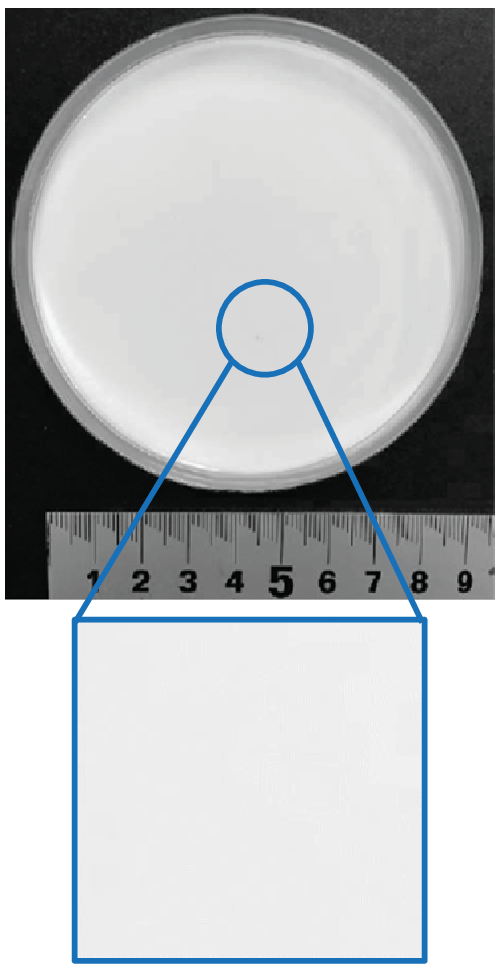}
\includegraphics[width=4.1cm]{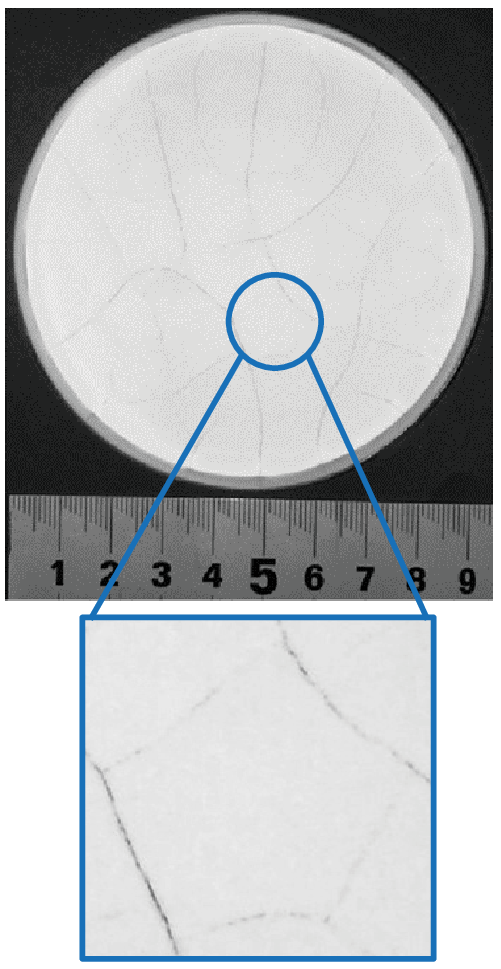}
\includegraphics[width=4.1cm]{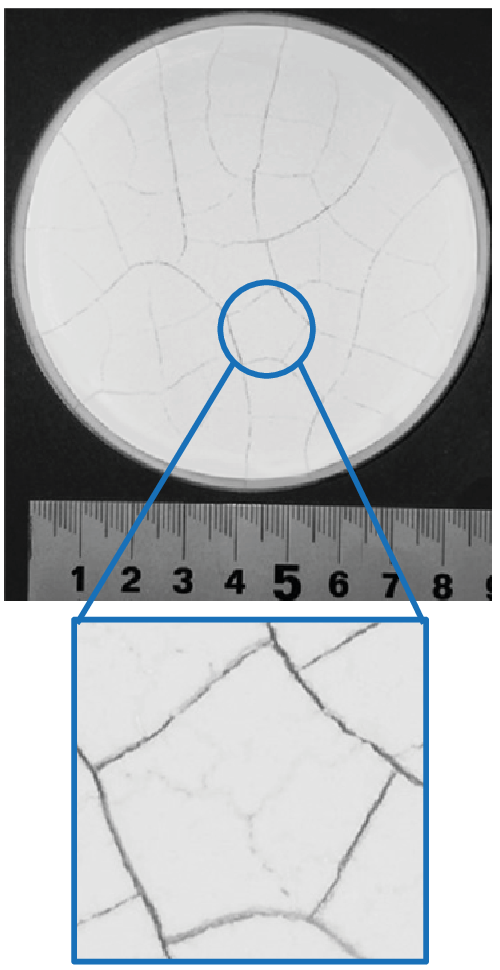}
\includegraphics[width=4.1cm]{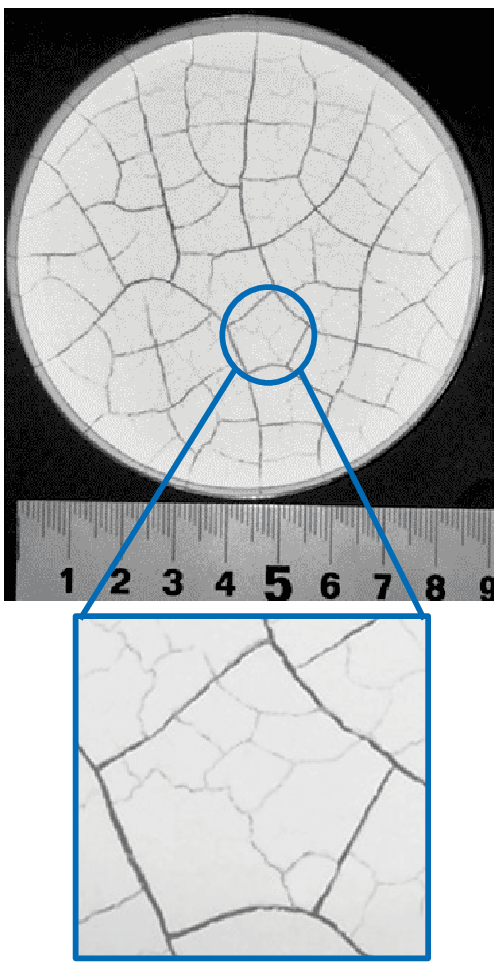}
\caption{
Corn starch slurries with 2-mm thickness
showing the sequential formation of a few primary cracks followed by the simultaneous formation
of many secondary cracks.
The time of drying is 30 min, 47 min, 89 min, and 276 min from left to right.
}
\label{fig:corncrackphoto}
\end{figure*}

Figure \ref{fig:particlesize}
shows the probability distribution of the grain size 
for the two kinds of starches
detected by the image-processing software 
(NI Vision Assistant 2012, National Instrument).
The horizontal axis indicates 
the equivalent circular diameter
of the starch grains,
{\it i.e.,} the diameter of a circular disk having the same area
as the cross-sectional area of the grain.
The number of grain samples that we measured
as well as the median of the 
equivalent circular diameter
that was evaluated
(marked by upward arrows)
are displayed in the legend of the figure.
From Fig.~\ref{fig:particlesize}, it follows
that potato starch shows a wide variation in size ranging from 6 $\mu$m
to 40 $\mu$m, which is in good agreement with the earlier study \cite{Costa2013},
and corn starch exhibits a narrow size distribution ranging from 4 $\mu$m to 14 $\mu$m.
In addition, the median for the grain size of potato starch ($=15.3$ $\mu$m)
is significantly larger than that of corn starch ($=6.35$ $\mu$m).

\section{Results}

\subsection{Time evolution of crack pattern}

Careful observation of the crack development during water evaporation
revealed a distinct difference in the crack morphology for potato and corn starches.
An important finding is the significant suppression of primary cracks in potato starch slurries.
Figure \ref{fig:potatocrackphoto} demonstrates
the cracking process of a potato starch slurry with a 2-mm-thick starch layer.
The slurry surface remained intact until 60 min after the start of drying.
Over time,
a multitude of secondary cracks (not primary ones) initiated simultaneously on the surface.
The number of secondary cracks and their intersections increased gradually as evaporation continued,
and the surface was split into many polygonal cells.
Once a portion of the slurry surface was split into sufficiently small cells,
no additional cracks formed inside the cells because a sufficient amount of tensile stress had already been released.
Eventually, the crack pattern saturated after 360 min,
at which the slurry thickness was reduced to 95 \% of the initial state
for all samples.

The absence of primary cracks for the case of potato starch
is in contrast with the case of corn starch,
where primary cracks dominate the early stage of desiccation crack development.
Figure \ref{fig:corncrackphoto} shows the results for a corn starch slurry with a 2-mm-thick layer.
At the initial stage (47 min after the start of drying), 
long primary cracks appeared one-by-one,
breaking the slurry surface in a sequential manner into several domains.
Every two connected primary cracks showed an almost vertical intersection,
as in the agreement with earlier studies on various kinds of 
water-grain mixtures \cite{Costa2013,DeCarlo2014,Khatum2015,Nandakishore2016}.
It is known that the spacing between primary cracks occurring on the dried corn starch surface
is several times larger than the starch layer thickness.
Afterwards, 
primary crack formation stopped ({\it e.g.}, 89 min after the start of drying in Fig.~\ref{fig:corncrackphoto}); instead, secondary cracks started to appear
inside the domains surrounded by the primary cracks.
At the final stage (276 min),
the dried slurry was finely divided by a combination 
of primary and secondary cracks;
for all samples, 
we observed the 5\% reduction of the slurry thickness
as was the case with potato starch.

Our experiments confirmed that the significant suppression of primary crack formation
in the potato starch slurries holds true even when we vary
the thickness of the slurry layer within the range of 2 mm to 11 mm,
or when we replace the container with those having a square shape with linear dimensions of 200 mm or less.
The only exception was the potato starch sample, which was poured into a very large square container with an area of 310$\times$230 mm$^2$
in area, which had an 80-mm-thick starch layer.
In the latter case, one primary crack was initiated from the edge of the container
and propagated toward the center, but it stopped without penetrating the full width
of the container.
These results demonstrated that there is a minor contribution of the homogeneous volume shrinkage,
which generally triggers the primary crack formation,
to the surface fracture of potato starch slurries.

\subsection{Drying curve}

\begin{figure}[ttt]
\includegraphics[width=8.5cm]{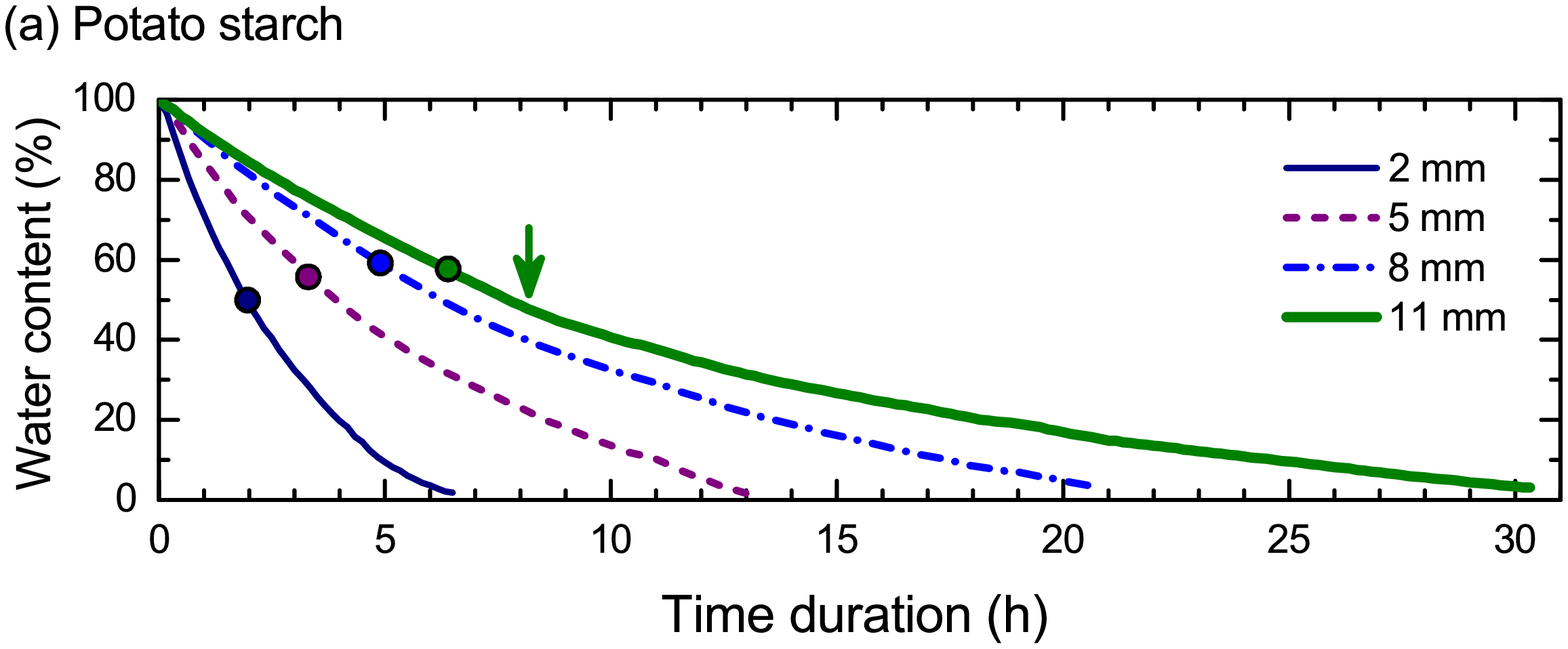}
\includegraphics[width=8.5cm]{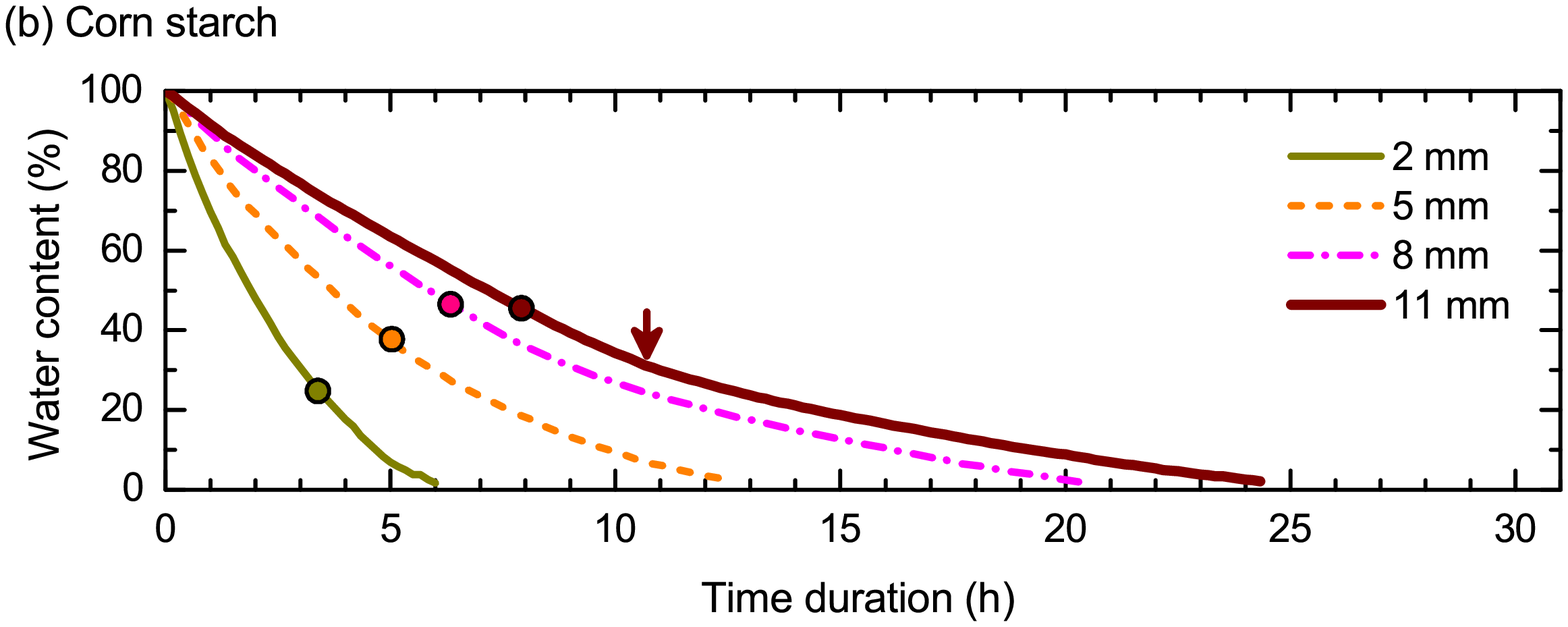}
\caption{
Time variation for the amount of water content in the starch slurries.
Solid circles placed on the curves indicate the point at which
secondary cracks were initiated.
Arrows indicate the turning point of the curve associated with the 11-mm-thick sample
across which the slope of the curve changes.
}
\label{fig:watercontent}
\end{figure}

Figure \ref{fig:watercontent} presents
the time variation with respect to the water content in the starch slurries.
For both potato and corn starches,
the slope of the curves at the initial drying stage ({\it e.g.}, $t\le 1$ h)
was found to decrease with increasing thickness of the slurry samples.
In addition, all curves were downward convex with respect to the time duration.
More specifically for each sample with a layer thickness of 5-11 mm,
there is a shallow turning point on the curve
(marked by arrows only for the 11-mm-thick sample)
across which the slope of the curve
changes significantly.
The drying time associated with the turning point serves as an indicator,
immediately before which secondary cracks started to form at the top surface of the slurries (marked by solid circles),
as was originally suggested by Ref.~[\onlinecite{Goehring2006}].
After the secondary crack initiation,
those cracks penetrate the depth over time,
during which the rate of water content evaporation slightly declines,
as seen in Figs.~\ref{fig:watercontent}(a) and \ref{fig:watercontent}(b).

\subsection{Geometry of crack pattern}

The geometry of the random polygonal tessellation that we obtained
was characterized from the perspective of the following three quantities:
the polygon order $N$
({\it i.e.,} the number of vertices involved in a given polygonal cell),
the polygonal cell's area $S$,
and the joint angle at the crack intersection.
For a given slurry layer thickness,
we prepared five different samples of completely dried slurries,
 and examined them to evaluate the probability distribution
of the three characteristic quantities listed above.
We confirmed that one slurry sample involved several tens or hundreds of polygonal cells.

\begin{figure*}[ttt]
\includegraphics[width=5.9cm]{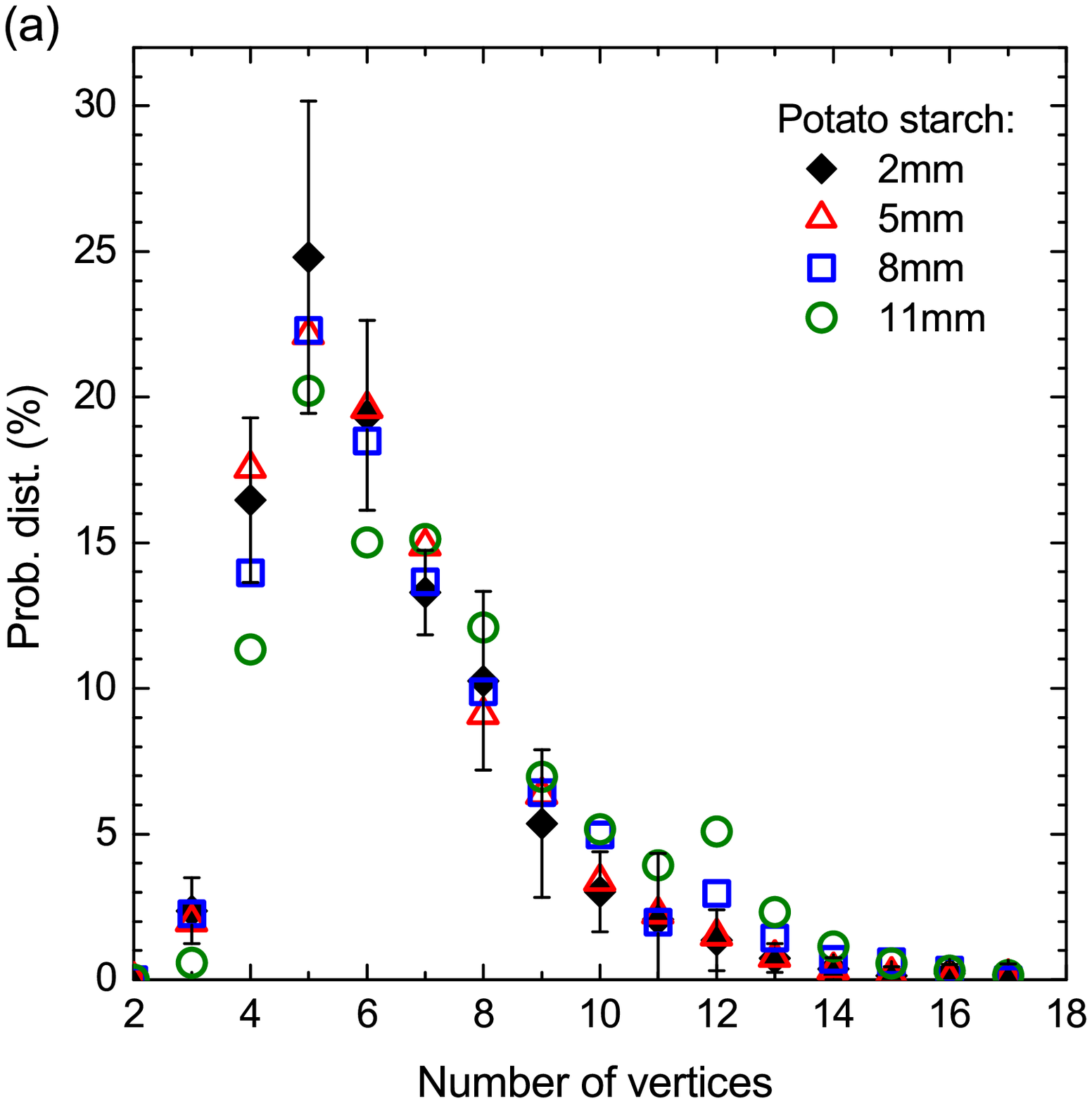}
\includegraphics[width=5.9cm]{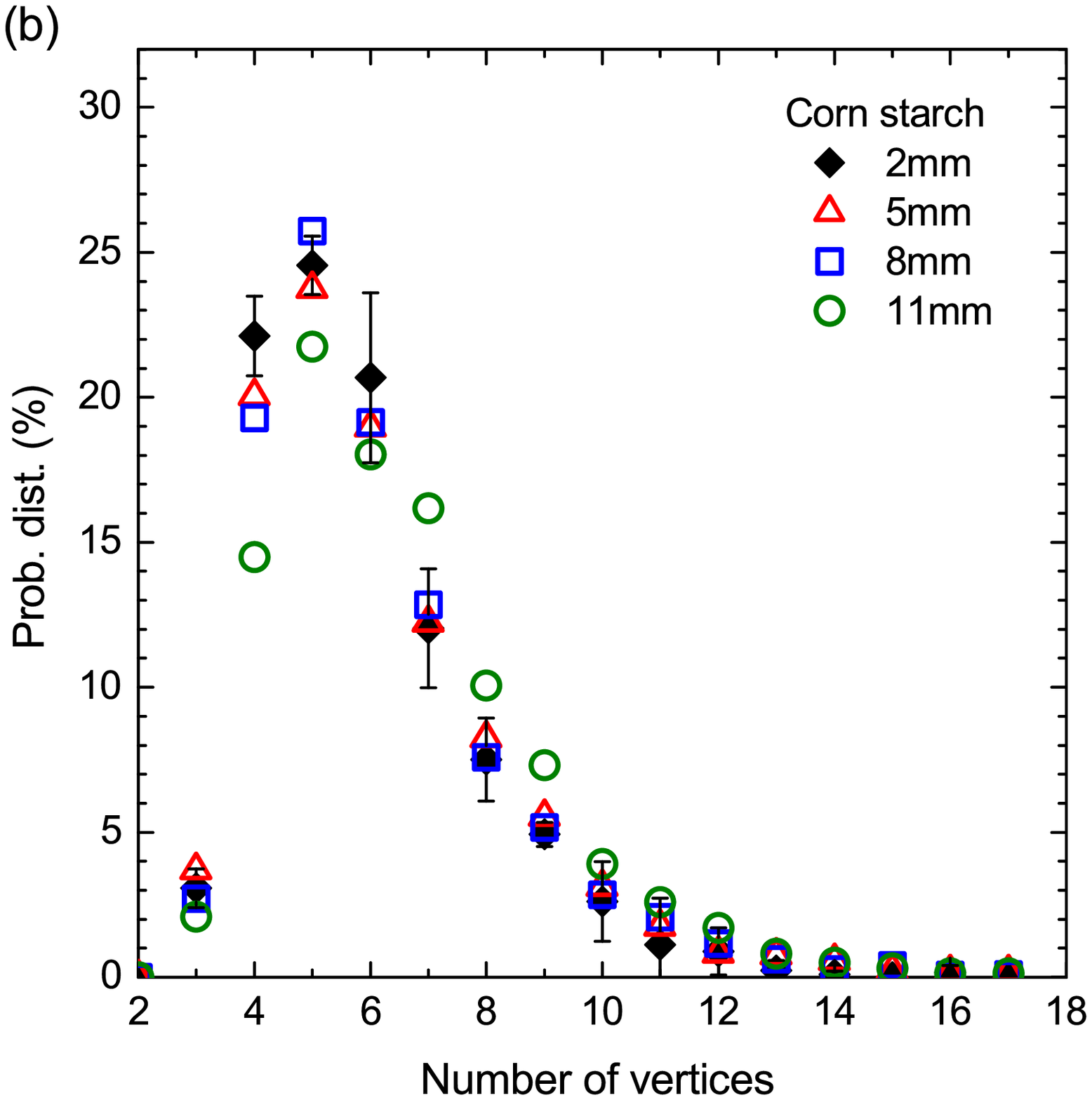}
\hspace*{12pt}
\includegraphics[width=3.6cm]{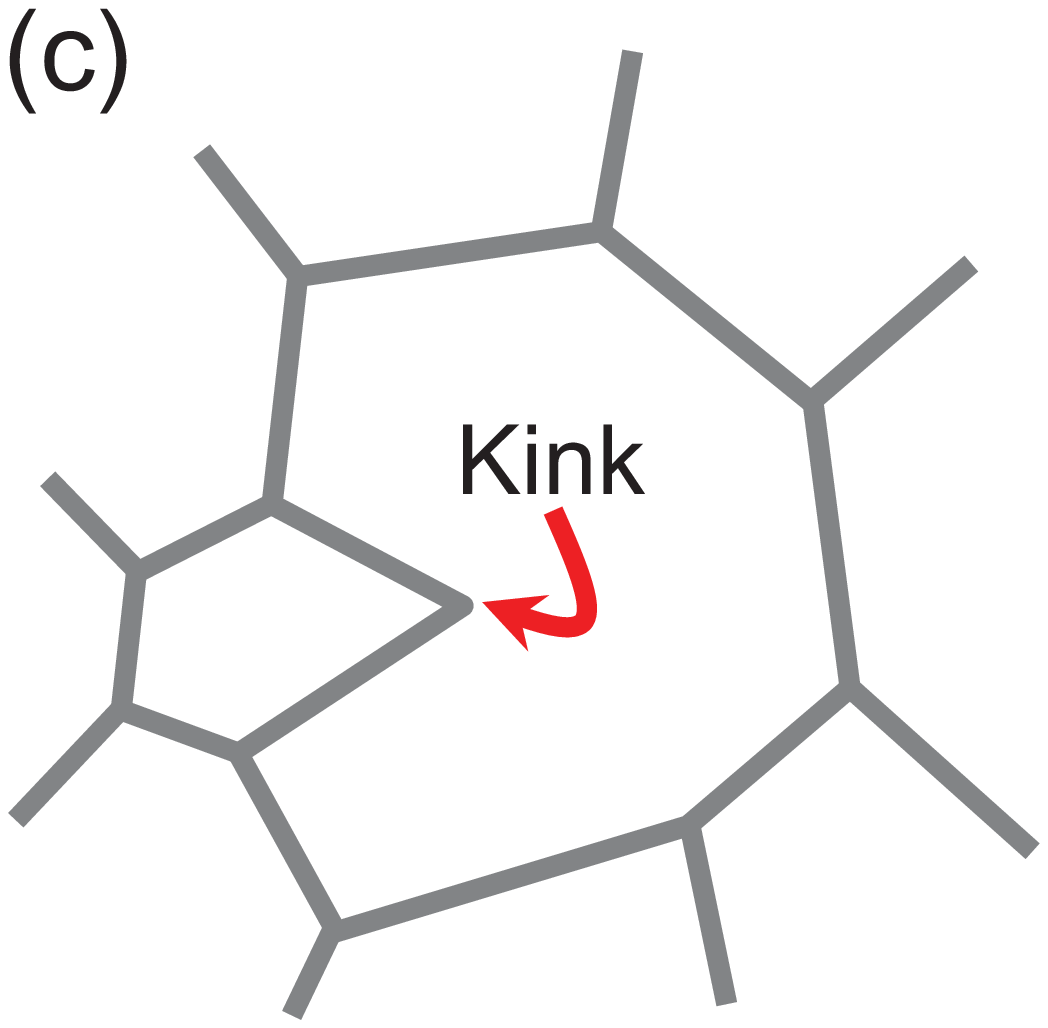}
\caption{
Probability distribution curve for the number of vertices $N$
involved in a given polygonal cell.
For each slurry thickness,
the averaged values over five different slurries samples are presented.
(a) Potato starch, (b) Corn starch,
(c) Schematics of an inward-convex polygonal cell having a kink.
}
\label{fig:vertex}
\end{figure*}

Figures \ref{fig:vertex}(a) and \ref{fig:vertex}(b)
show the probability distribution of the polygon order $N$.
The two graphs clearly show
an overwhelming majority of pentagonal cells ($N=5$),
together with squares ($N=4$) and hexagons ($N=6$).
We found that the distribution curves shift slightly 
to the right
({\it i.e.,} to the larger $N$ region)
as the layer thickness increases from 2 mm to 11 mm,
while keeping the peak position at $N=5$.
The dominancy of pentagons agrees in part with the previous result on potato starch samples \cite{Toramaru2004}; yet it seems in contrast with those reported by Ref.~[\onlinecite{Muller1998}], which showed that hexagons dominated in corn starch slurries that were much thicker than those we have used. This issue will be revisited in Section IV-B.
We also found in Figs.~\ref{fig:vertex}(a) and \ref{fig:vertex}(b) the
long tail in the distribution curve
for large $N$ ({\it e.g.,} $N\ge 8$),which
reflects the presence of many polygonal cells with kinks along their edges; 
see Fig.~\ref{fig:vertex}(c).
The kinks, from which only two cracks branch off,
cause the adjacent polygonal cells to be inward convex;
thus, they tend to increase the value of $N$ for those cells.

\begin{figure*}[ttt]
\includegraphics[width=5.9cm]{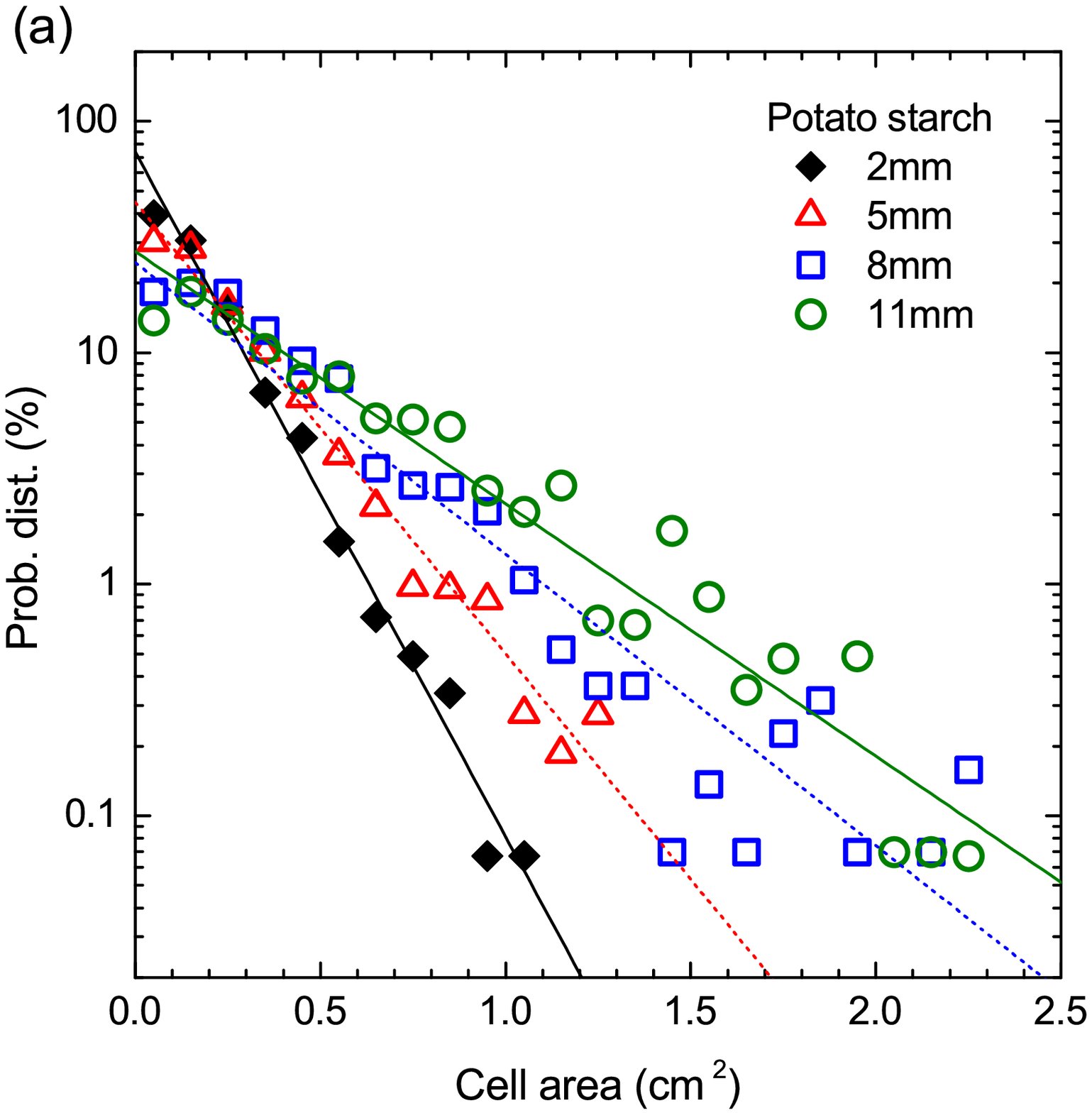}
\includegraphics[width=5.9cm]{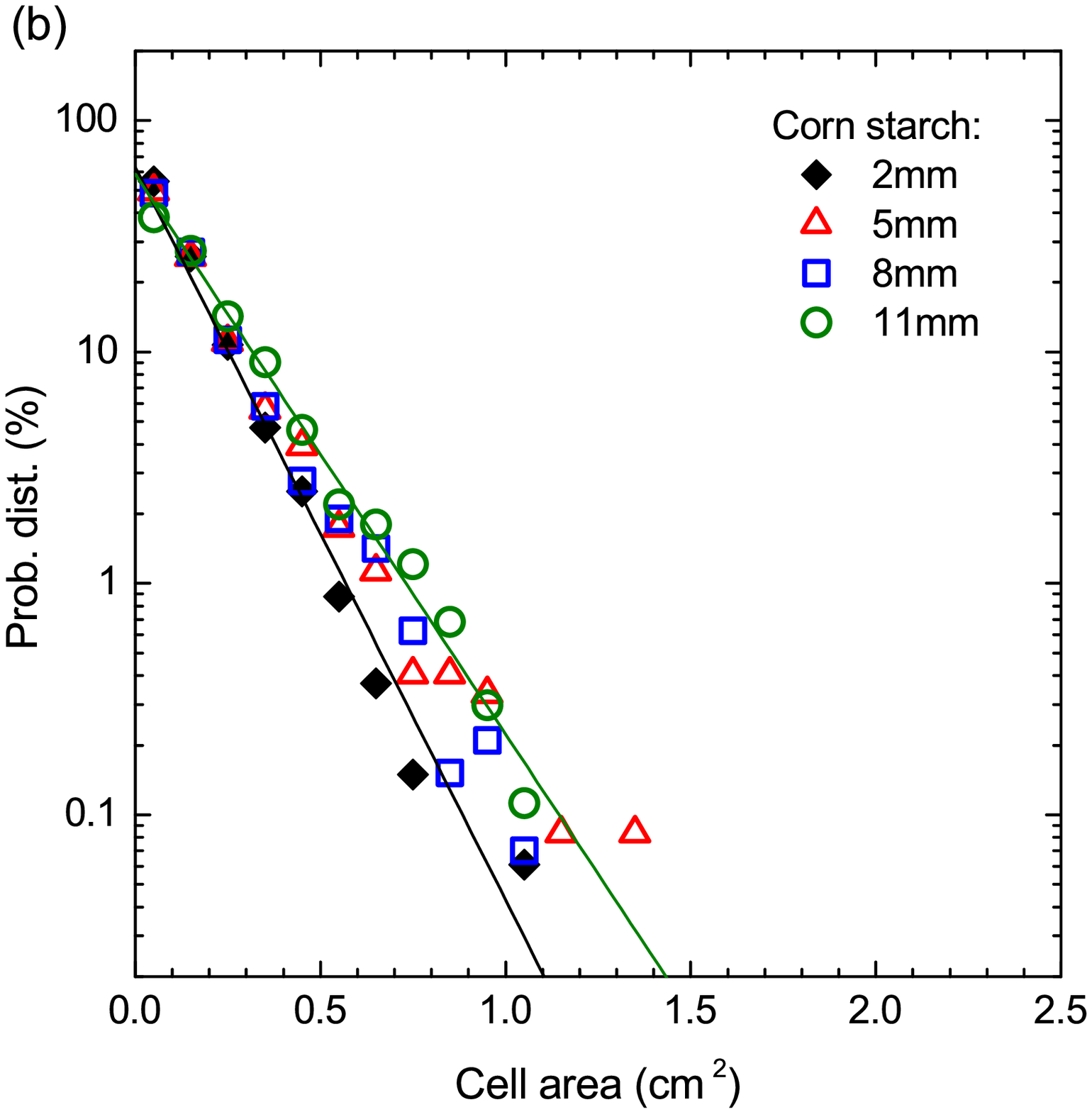}
\hspace*{12pt}
\includegraphics[width=4.2cm]{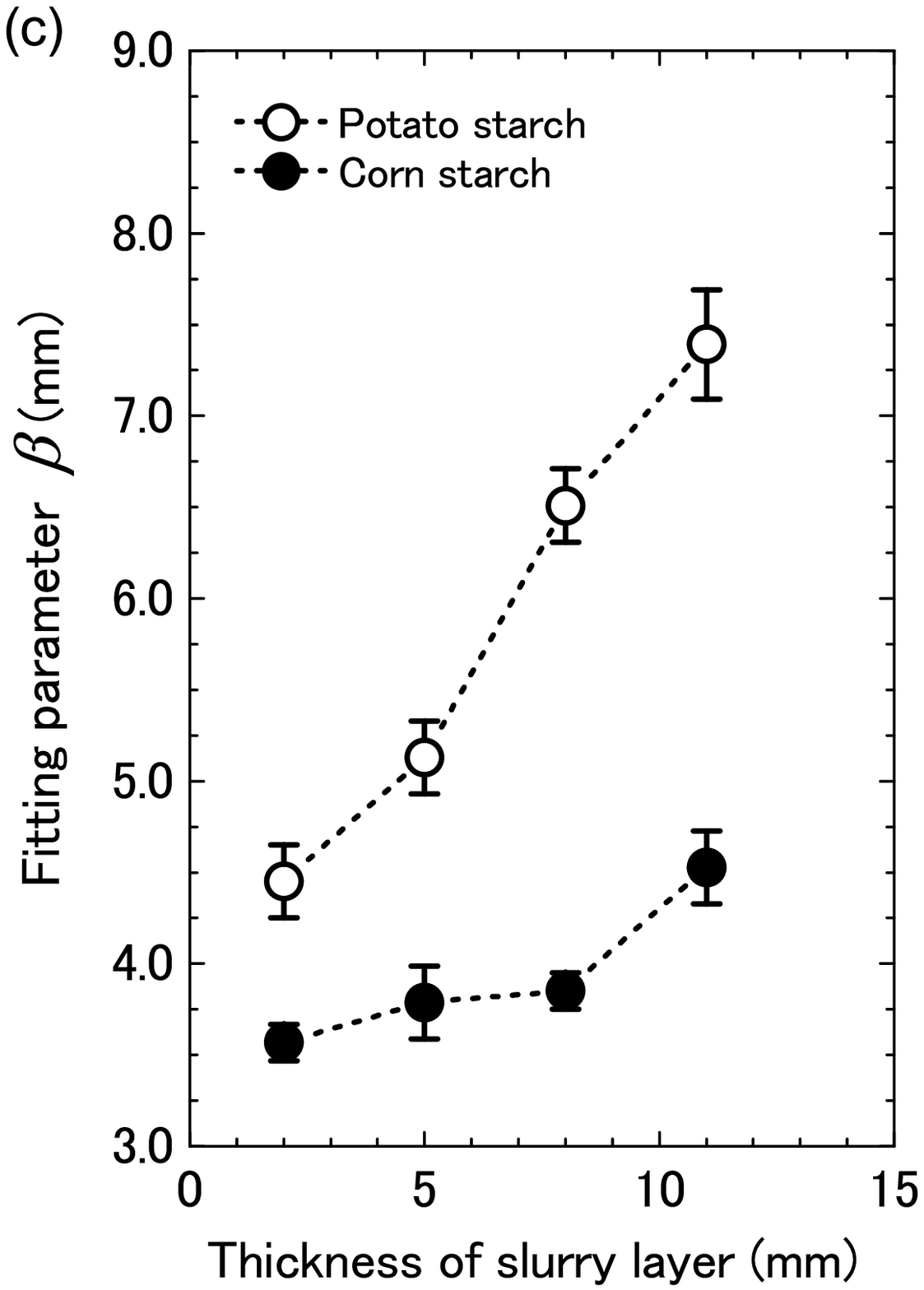}
\caption{
Probability distribution curve of the area $S$ of polygonal cells:
(a) Potato starch, and (b) Corn starch.
For each value of slurry thickness,
we present the averaged values for five different slurries samples.
For potato starch, the figure shows the exponential fitting curves for samples with all thicknesses. For corn starch, it shows the curves for samples with thicknesses of only 2mm and 11mm.
(c) Thickness dependence of the fitting parameter $\beta$ defined by Eq.~(\ref{eq_003}).}
\label{fig:area}
\end{figure*}

Figures \ref{fig:area}(a) and \ref{fig:area}(b) illustrate
the probability distribution of the area of polygonal cells surrounded by secondary cracks
using a semi-logarithmic scale.
From Fig.~\ref{fig:area}(a), it follows that
a thicker slurry of potato starch tends to provide
more polygonal cells over a wide area.
In the case of corn starch, we found no such thickness effect on the polygonal cell area,
as demonstrated in Fig.~\ref{fig:area}(b);
in the latter case,
a thickening of starch slurries
caused only a slight shift of the data points in the plot to the right.
To quantify the effect of the layer thickness,
we fitted the data points onto an exponential curve defined by
\begin{equation}
P(S) = c \cdot \exp\left( - \frac{S}{\beta^2} \right), \label{eq_003}
\end{equation}
and we evaluated the fitting parameter $\beta$ (as well as the coefficient $c$)
as a function of the layer thickness $L_{\rm th}$.
The parameter $\beta$ characterizes the decay length of internal stress away from a crack in a thin layer \cite{Beuth1992,ZCXia2000}.
Figure \ref{fig:area}(c) shows that for potato starch,
$\beta$ increases significantly with increasing $L_{\rm th}$
in an almost linear manner from 
$\beta\sim 4.4 \pm 0.2$ mm 
at $L_{\rm th} = 2$ mm
to 
$\beta \sim 7.4 \pm 0.3$ mm 
at $L_{\rm th} = 11$ mm.
For corn starch, the increase rate of $\beta$ is subtle
by comparison with the case of potato starch;
it varies from 
$\beta \sim 3.6 \pm 0.1$ mm 
to 
$\beta \sim 4.5 \pm 0.2$ mm 
with increasing $L_{\rm th}$.

\begin{figure}[ttt]
\includegraphics[width=5.9cm]{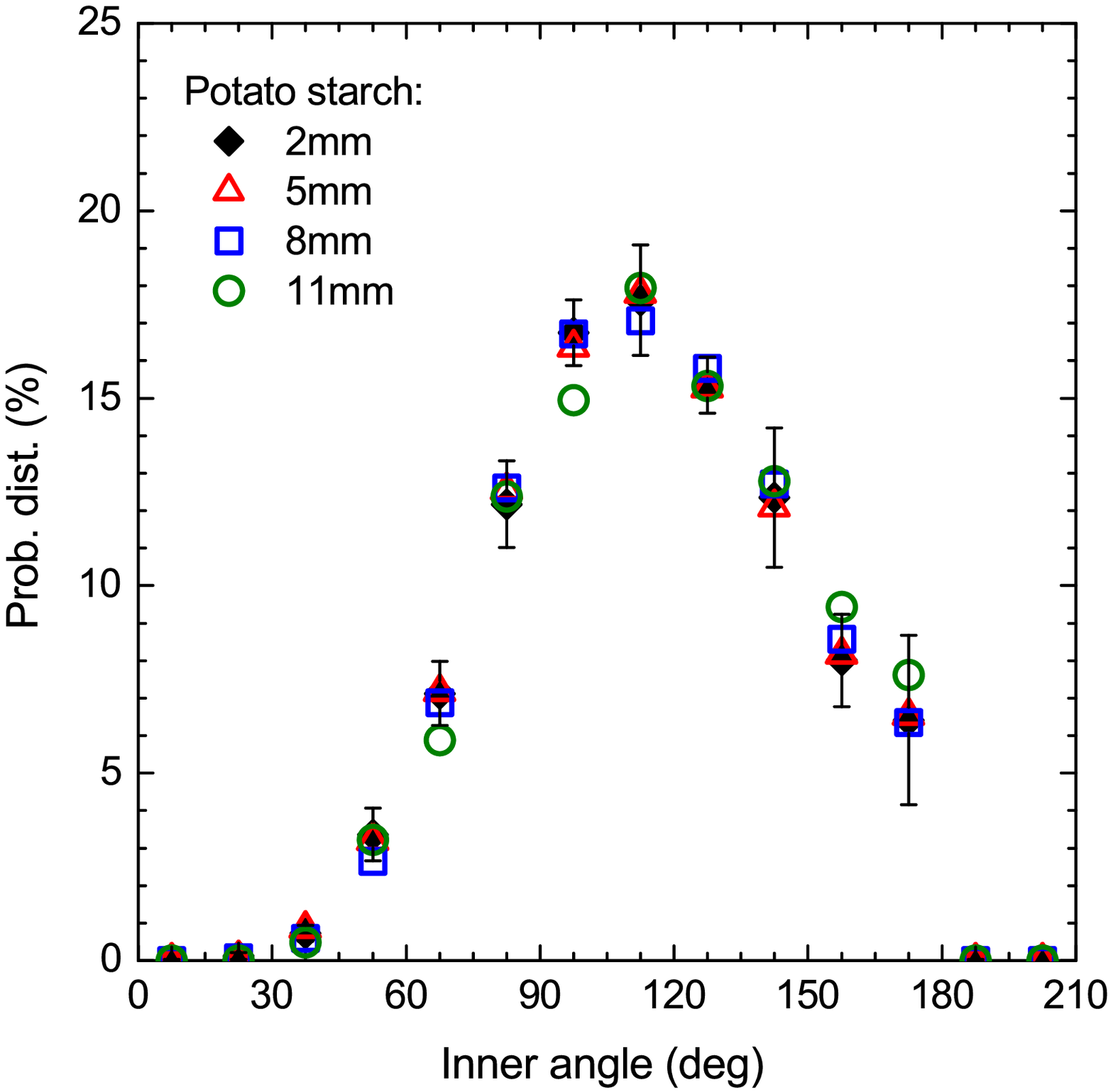}
\includegraphics[width=5.9cm]{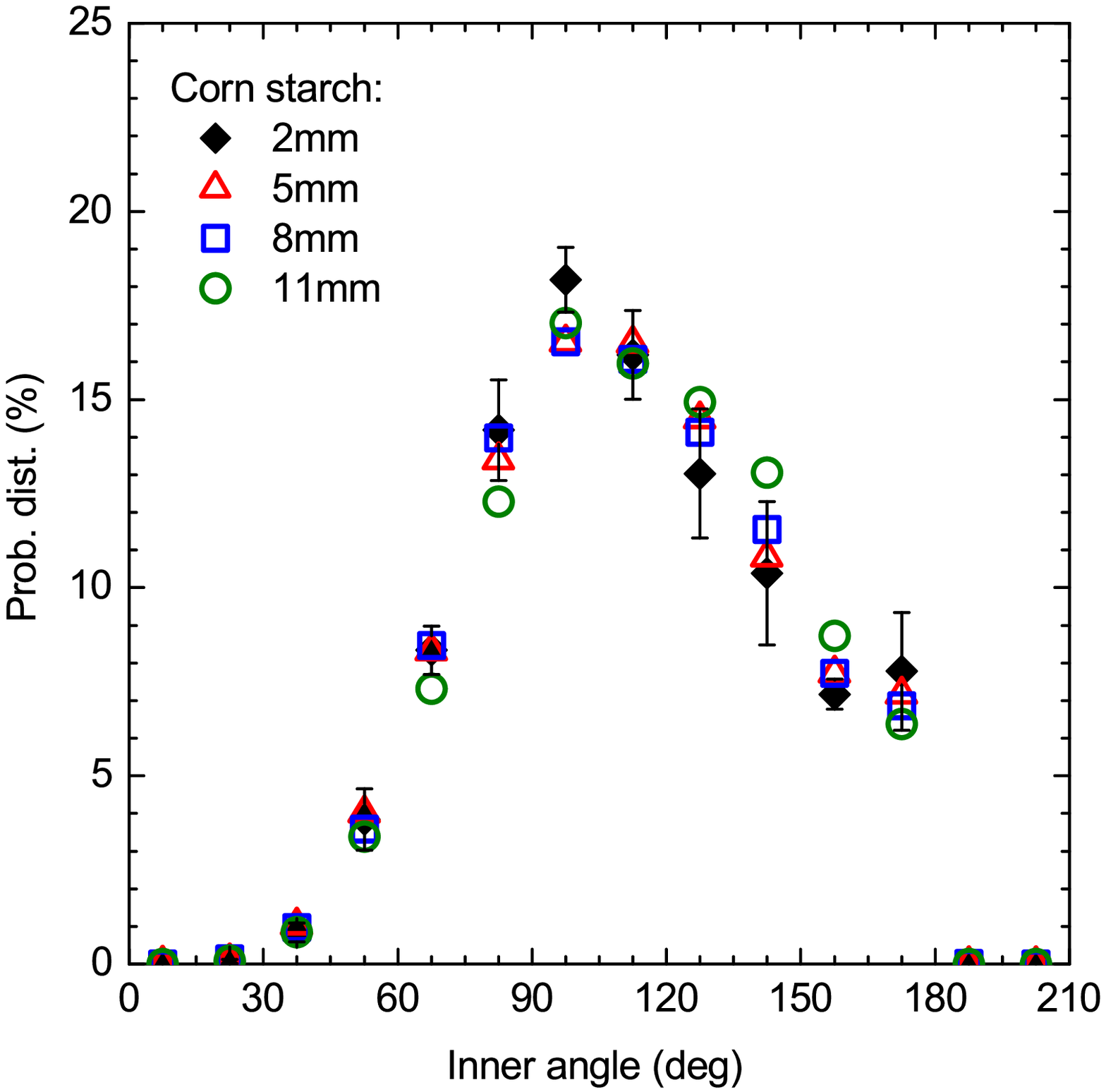}
\caption{
Probability distribution curve of the intersection angle between cracks.
(a) Potato starch. The peak is located within the range 105$^\circ$-120$^\circ$.
(b) Corn starch. The peak is located within the range 90$^\circ$-105$^\circ$.
}
\label{fig:angle}
\end{figure}

Figure \ref{fig:angle} shows the distribution of the intersection angle between cracks.
The two plots show that there is no significant difference in the distribution curve
between the two starch ingredients,
nor is there a dependence on the layer thickness.
Nevertheless,
the peak position of the distribution curve
differs sufficiently between the two starch ingredients.
In other words,
potato starch shows a peak at the range of 105$^\circ$-120$^\circ$,
while corn starch shows a peak at 90$^\circ$-105$^\circ$.

It should be remarked that 
the peak position for corn starch samples does not coincide with
that reported earlier;
it was shown in Ref.~[\onlinecite{Mizuguchi2005}] that
the peak located at 120$^\circ$-130$^\circ$ for corn starch
and all the measurement data of joint angles 
fell into the range from 90$^\circ$ to 180$^\circ$.
As well, the standard deviation of the joint angles, $\sigma_{\theta}$,
estimated by the present analysis shows a certain deviation
from the earlier one; it was reported in Ref.~[\onlinecite{Goehring2005}] that
$\sigma_{\theta} \sim 25^\circ$ for corn starch samples,
which is smaller than our result of $\sigma_{\theta} \sim 33^\circ$.
These differences in the probability distribution of joint angles for corn starch may be attributed to the difference either in experimental conditions or 
the methodology for quantifying joint angles, 
though the detailed understanding has not yet been reached. 
Nevertheless, the singleness of the peak in the joint angle distribution 
was observed in common for all experiments,
indicating the presence of a frequently occurring joint angle 
under a given experimental condition.

\section{Discussion}

\subsection{Spatio-temporal variation in local water content}

\begin{figure}[ttt]
\includegraphics[width=7.0cm]{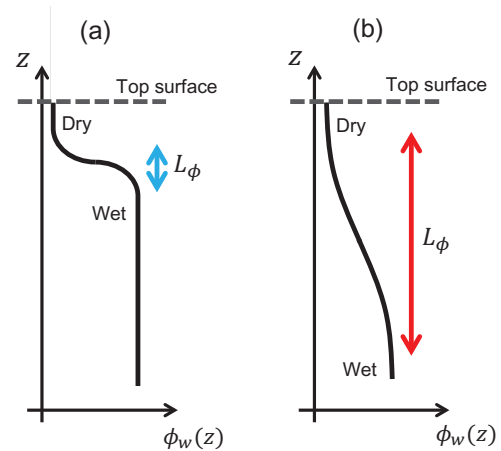}
\caption{
Schematic illustration of the variation in the water volume fraction $\phi_w(z)$
in the vertical direction labeled by $z$.
$L_{\phi}$ indicates the vertical limits showing the gradient of $\phi_w(z)$
within which the tensile stress is concentrated.
(a) The case of low diffusivity $D(\phi_w)$ around the intermediate zone.
(b) The case of high diffusivity $D(\phi_w)$
}
\label{fig:dryfront}
\end{figure}

We observed that the potato starch slurries showed a systematic dependence 
of the cell area distribution $P(S)$ on the slurry thickness.
In contrast, for corn starch, the thickness of the slurry layers
had no significant effect on the curve of $P(S)$.
This contrasting behavior can be explained by considering
the spatio-temporal variation in the local volume fraction of water, $\phi_w(z,t)$,
with $z$ being the vertical axis.
In the initial case, the starch slurries are saturated by liquid water, 
and thus $\phi_w$ is constant, as designated by $\phi_w^{\rm max}$,
over the whole precipitated starch.
As evaporation continues, $\phi_w$ near to the top, air-exposed surface
decreases with time, while the corresponding value far below the top surface remains at the initial value.
This gives rise to the development of an intermediate zone 
that is sandwiched by the top drying surface and the far-below fully wet slurry,
through which $\phi_w(z)$ undergoes a gradient with respect to $z$ from $\phi_w \simeq 0$ at the top surface
to $\phi_w=\phi_w^{\rm max}$ at the bottom \cite{Mizuguchi2005,GoehringPRE2009}.
The vertical limits of the intermediate zone depends on the magnitude of the local diffusivity $D(\phi_w)$
of the water content in the slurry.
It was proposed in Ref.~[\onlinecite{GoehringPRE2009}] that
both the capillary and vapor transport of water can be described
by an effective diffusion equation with a nonlinear diffusivity:
\begin{equation}
\frac{\pa \phi_w}{\pa t} = \frac{\pa}{\pa z} \left[ D(\phi_w) \frac{\pa \phi_w}{\pa z}  \right].
\label{eq_005}
\end{equation}
If the value of $D$ around the intermediate zone is sufficiently large, 
the water content that had already evaporated from the top surface 
will be compensated by water diffusion from the wetter portion of the slurry far below the top (very dried) surface
via transportation through capillary bridges.
This water supply from bottom to top causes an expansion of the vertical limits of
the intermediate zone,
reducing the slope of $\phi_w(z)$ with respect to $z$ within the zone,
as illustrated in Fig.~\ref{fig:dryfront}.
On the contrary, if the value of $D$ around the intermediate zone is sufficiently small,
the water content that is evaporated is not compensated from the bottom,
so the stepwise profile of $\phi_w(z)$ as a function of $z$
does not change considerably.
It is natural to assume that 
$D$ should be large for the slurry consisting of large-sized, irregular-shaped grains,
because there is a large available pore space through which water can diffuse.
It is thus inferred that for the potato starch slurries,
the vertical limits that indicate the $\phi_w(z)$ gradient
becomes enlarged
compared to the corresponding values for corn starch slurries.

It is important that the inhomogeneous tensile stresses that cause secondary cracking are
concentrated within the intermediate zone in which $\phi_w(z)$ shows a gradient variation.
This is because in the slurry portion far below the air-exposed surface,
there exists a sufficient amount of water; therefore, capillary transport smooths out any spatial
variation in the local volume fraction of water.
This smoothing effect causes the inhomogeneity in the capillary-induced tensile stress to be eliminated.
In this context, a potato starch slurry involves a ``thick" intermediate zone
in which tensile stress is concentrated.
In contrast, a corn starch slurry involves a ``thin" intermediate zone
in the same sense.
Below the stress-concentrated intermediate zone having a finite thickness,
the interior of the slurry is more moist, so it can be considered as a different medium
trying to hold the upper dried layer together and providing adhesion.
Under this condition,
the intercrack spacing on the two dried surfaces
should be proportional to the thickness of the layer 
under tensile stress \cite{Beuth1992,ZCXia2000},
as a direct consequence of the theory of fracture mechanics;
the validity of the proportionality relation 
has been experimentally confirmed
for many kinds of desiccated slurries
\cite{Groisman1994,Shorlin2000,Colina2000EPJE,BohnPRE_From_2005}.
Eventually, the vertical limits of the stress-concentrated region determine
the horizontal spacing between adjacent secondary cracks,
or equivalently,
they determine the area of the polygonal cells surrounded by secondary cracks

Our experimental results indicate that for the potato starch,
the vertical limits of the stress-concentrated region, which is denoted by $L_\phi$,
is comparable to or exceeds 11 mm,
which is the maximum of the slurry thickness that we used in the experiments.
Accordingly, $L_\phi$ was truncated by the slurry thickness $L_{\rm th} (\le L_\phi)$
for all the samples.
As a result, $L_{\rm th}$ serves effectively as $L_\phi$,
and determines the crack spacing instead of $L_\phi$.
This is the reason 
why the cell area distribution $P(S)$ showed a strong dependence on the slurry thickness.
The larger $L_{\rm th}$ leads to the tendency of wider polygonal cell formation
in potato starch samples,
as characterized by the steep increase
in the fitting parameter $\beta$
with increasing $L_{\rm th}$ [see Fig.~\ref{fig:area}(c)].
In contrast, $L_\phi$ for the corn starch is believed to be less than 11 mm,
and as a result, $P(S)$ did not show 
significant dependence on $L_{\rm th}$.
Our speculation of $L_{\rm \phi} \le 11$ mm for corn starch
is consistent with the existing studies based on X-ray tomography \cite{Mizuguchi2005} 
and destructive sampling \cite{GoehringPRE2009},
which unveiled the presence of a sharp dry front 
with a thickness of a few millimeters.

\begin{figure}[ttt]
\includegraphics[width=8.0cm]{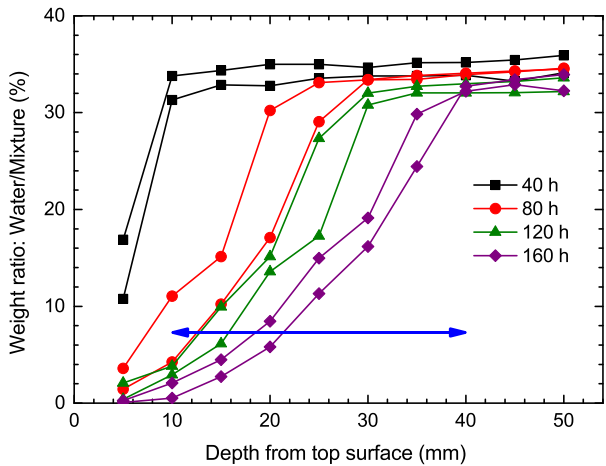}
\caption{
Water content distribution in 50 mm-thick potato starch slurries.
Four pairs from the eight slurry samples are dried for different time periods 
ranging from 40 hours to 160 hours.
The horizontal array indicates the intermediate zone
with $L_{\phi}$ in thickness.
}
\label{fig:moisture}
\end{figure}

To further examine the validity of the estimation that 
$L_{\phi}$ for potato starch should be larger than that for corn starch,
we have measured the spatial distribution of the water content
within potato starch samples with 50 mm in slurry thickness;
see Appendix B for the method of measurement.
Figure \ref{fig:moisture} demonstrates the drying time dependence of 
the water distribution within the potato starch slurries;
the horizontal arrow indicates the intermediate zone
with $L_{\phi}$ in thickness.
It follows from the figure that $L_{\phi}$ for the potato starch is 
estimated to be ca. 30 mm,
which provides an evidence that $L_{\phi}$ for potato starch
should be larger than 11 mm as described above.

\subsection{Predominance of pentagonal cells}

Another important fact that was deduced from the 
present work is that
the peak in the distribution of the polygon order $N$
is at $N=5$ both for potato and corn starches.
This may appear to be counterintuitive, because 
it is often said that 
hexagonally ordered patterns are found dominantly 
in many settings \cite{Goehring2013}.
We conjecture
that the predominance of pentagons that we observed
originates from the 
relatively high rate of desiccation,
compared to the rate under natural drying condition;
the relatively high desiccation rate was 
a consequence of heat supply 
from the lamp set up above the slurry surface.
This conjecture is consistent with the field observation of columnar joints
made from the cooling lava flow,
as briefly discussed below.

Columnar jointing is a spectacular geological structure, 
composed of an array of prismatic columns
that shows strong regularity in the cross section.
The regularity in the polygonal cracking network 
in the cross section is a result of
inward penetration of thermal contraction cracks 
that form at the surface of cooled lava.
Though the surface cracks are not so regular as those 
in the cross section,
interaction of adjacent cracks during inward penetration
is believed to lead to changes
in the spacing between cracks
and in the orientation of crack development,
then resulting in the long-range ordered 
polygonal pattern at the cross section
inside the cooled lava \cite{Degraff1993,Budkewitsch1994,Hofmann2015}.
The occurrence frequency of the polygon order $N$
in columnar jointed rocks worldwide 
has been summarized in Ref.~[\onlinecite{Budkewitsch1994}].
Attention should be paid to the fact that
the hexagon is not necessarily preferred in columnar joint formation.
The predominance of pentagons is exemplified by Burntisland dyke in Scotland \cite{Sosman1916} and 
Mt. Rodeix basalt in France \cite{Beard1959},
whereas Giant's Causeway in Ireland \cite{OReilly1879} exhibits a larger number of hexagons.
The difference in the preferred type of polygons
is known to be due to the difference in the cooling rate of lava.
The cooling starts from the top 
and bottom surfaces of the pooled lava,
generating the variation in the internal temperature at different depths \cite{Goehring2008,GoehringPNAS2009}.
This is similar to the case of desiccation cracking, where
the drying generates a variation in the water volume fraction.
In the cooled lava,
thermal shrinkage provides the driving force for the crack propagation at different depths,
engendering the highly ordered array of prisms with a polygonal cross section.

It should be noted that for columnar joints,
the dominant polygon order $N$ depends on the cooling rate \cite{Toramaru2004}.
More precisely, 
it was proposed that a
pentagon be preferred as the columnar shape at a higher cooling rate,
whereas a hexagon is preferred at a lower cooling rate.
From the analogy in the formation mechanism 
between the columnar joints and dried starch cracking \cite{GoehringPNAS2009},
it is therefore natural to deduce that 
a pentagon is preferred in the polygonal cracking of dried starch
when the rate of drying is sufficiently large.
This conclusion is consistent qualitatively with the starch-water experiment
reported by Ref.~[\onlinecite{Toramaru2004}],
despite there being differences in the experimental conditions.

\subsection{Distribution of the intersection angles}

It is also interesting to note the persistence in the distribution profile
of intersection angles, regardless of the starch ingredient and the slurry thickness,
as demonstrated in Fig.~\ref{fig:angle}.
The peak position for the corn starch 
is located within the range of 90$^\circ$-105$^\circ$, which is slightly less than that for potato starch. This decrease
is attributed to the presence of primary cracks
that propagate linearly through the slurry surface
from one side edge of the slurry to the other side edge,
or to the other existing primary crack.
The intersection angles between a primary crack and a secondary crack,
as well as those between two primary cracks,
are likely to be perpendicular because of the crack-induced stress-release mechanism.
Near an edge of the medium or near an existing primary crack,
the stress occurs only in the direction parallel to the edge. 
Therefore, 
cracks should propagate normal to the direction along which the stress is highest
in order to produce the maximal relief of the stress.
This explains the peak of the angle distribution 
at slightly more than 90$^\circ$.
In the case of potato starch,
no primary crack was formed, so the random polygonal tessellation is totally composed
of secondary cracks.
Therefore, three-pronged junctions between cracks are dominant,
resulting in a majority of intersection angles
that are approximately 120$^\circ$.

The nucleation of Y-shaped cracks is also responsible for the frequent occurrence of the joint angles larger than 90$^\circ$. In real slurry samples, the air-exposed surface of a starch slurry involves a certain number of point-like defects, at which the cohesion between adjacent particles should be weak. These defects often give rise to the nucleation of three (or more) cracks that emanate from the point-like defects. The cracks joint by an angle larger than 90$^\circ$ from the beginning; thus, the development of these cracks contribute to the shift in the joint angle distribution toward the larger angle.

\section{Summary}

In this study, we experimentally studied the impact of the microscopic geometry of starch grains on the macroscopic patterns of polygonal cracking that occurs at the air-exposed surface of dried starch slurries. The difference in the size and shape of grains for potato and corn starches was responsible for the contrasting cracking patterns. Specifically, dried potato-starch slurries showed remarkable suppression of primary crack formation and a strong dependence of the cell-area distribution on the slurry thickness, as manifested by the large-sized, oval-shaped geometry of the constituent starch grains. The results can be explained by considering water diffusion around the intermediate zone, which is sandwiched by the dried upper surface and the moist bottom portion of the slurry under desiccation. The water diffusion is controlled by the microscopic size and shape of grains, determining the vertical limits of the stress-concentrated zone that dictates the macroscopic cracking patterns on the slurry. The predominance of pentagonal cells is attributed to the relatively high drying rate of the starch under the present condition, as a reminiscence of pentagon-dominated columnar joints that typically occur in the fast-cooling lava flow.

\section*{Acknowledgement}

We 
would like to express great appreciation to
Akio Nakahara, So Kitsunezaki,
Kenji Oguni, and Sayako Hirobe
for the fruitful discussions and valuable suggestions at the initial stage
of the present work.
We are also indebted to Akio Inoue for use of 
time lapse cameras as well as
helpful comments regarding the image analyses.
Our great thanks are also extended to Futaba Kazama and Kei Nishida 
for use of an incubator and a drying oven, respectively,
and Yasutada Suzuki for use of an infrared thermometer.
This work was supported by JSPS KAKENHI Grant Number JP25390147.

\appendix
\section*{Appendix A: Image processing workflow}

\begin{figure*}[ttt]
\includegraphics[width=16.0cm]{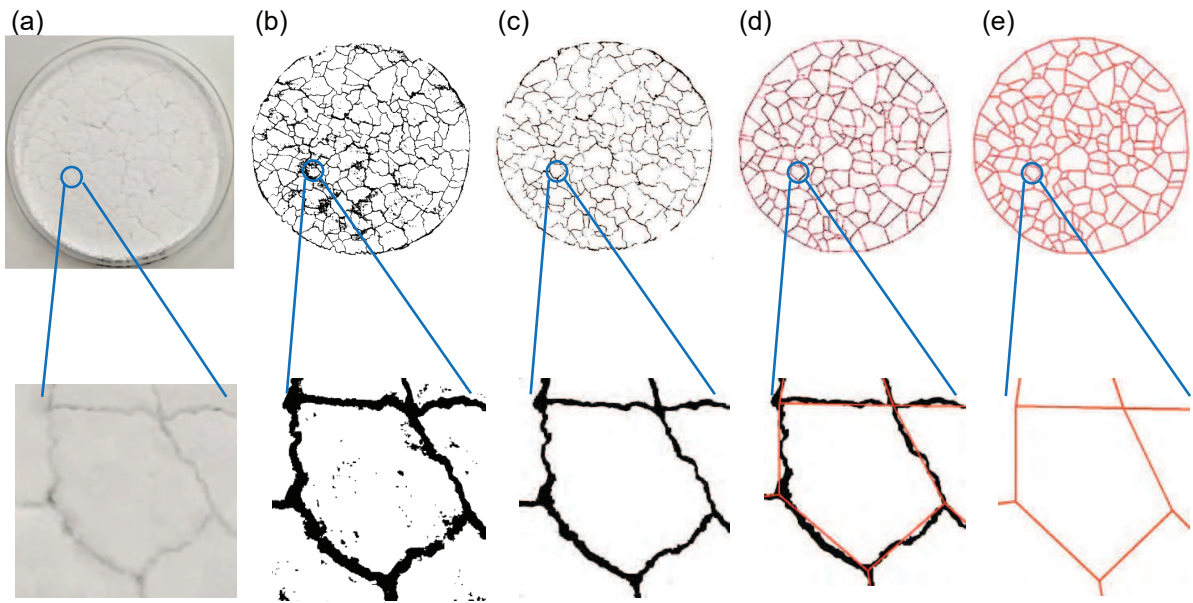}
\caption{
Diagram for the image processing.
(a) Raw picture of the fractured surface of a 5mm-thick potato starch slurry. 
(b) Raster image of the photo with enhancement in contrast ratio..
(c) Raster image after cleanup procedure.
(d) Superimposition of the vector image produced by ArcGIS (colored in red) with the cleanup raster image.
(e) Final image of the polygonal tessellation.
}
\label{fig:GISprocessing}
\end{figure*}

\vspace{12pt}

Figure \ref{fig:GISprocessing} shows the actual step of the image processing.
First, we prepared a raw picture of the air-exposed fractured surface 
obtained by complete drying of, for instance, a 5mm-thick potato starch slurry
as shown in the panel (a).
Next, we converted the picture into a raster image as shown in the panel (b).
The raster image was made up of square 
pixels
that are regularly spaced;
it was obtained by enhancing the contrast ratio of the scanned raw picture
followed by binary thresholding based on the Otsu algorithm \cite{Otsu1979}.

As a next step, we edited the raster image by filling the gaps 
that are a result of the poor quality of the scanned picture. 
In addition, we eliminated the unwanted 
pixels
that resulted from 
contrast enhancement. 
At this stage, we also erased the following two classes of 
pixels:
i) those corresponding to tiny cracks that were isolated from the main network,
and ii) those that branched off the network but were terminated 
without connecting to any other cracks. 
Through the editing, we prepared the cleanup raster image 
as shown in the panel (c).

As the final step, we used ArcGIS software \cite{arcgis}
to vectorize the cleanup raster image;
the vector image obtained is shown in the panel (d)
together with the cleanup raster image prior to vectorization.
In the raster-to-vector conversion, 
we needed to set several control parameters 
in order to choose the raster data to be vectorized and 
to determine how the geometry of the output vector data should be constructed. 
The optimal values of the control parameters for vectorization 
are not uniquely determined but depend on situations; 
we thus explored the appropriate settings by hand, 
and eventually obtained the vector image demonstrated in the panel (e).

\section*{Appendix B: Measurement of water content distribution in potato starch}

In the measurement, we prepared the 50 mm-thick potato starch slurries 
with 75 mm in diameter.
Eight slurry samples having the same size were prepared 
and dried beneath the 36-W lamp 
under the same conditions as described in Section II-A.
In order to evaluate the drying time dependence of 
the water distribution within the slurries,
we have assigned different drying time periods
({\it i.e.}, from 40 hours to 160 hours) 
to four pairs from the eight samples, 
as shown in the legend of Fig.~\ref{fig:moisture}.
After the drying time periods,
the dried slurries were cut into 10 round slices 
with equal thickness of 5 mm, and the weights of the slices were measured.
We then put the slices in a drying oven (DX402, Yamato Scientific)
with a constant temperature 
of 45 $^\circ$C and wait more than a whole day to completely remove 
the amount of water from the slices. 
By measuring the loss in weight of the slices, 
we have determined the amount of water that had been contained 
in the slices immediately after the drying time periods.
Eventually, we have obtained 
the spatial distribution of water content 
in the original 50 mm-thick potato starch slurries
and its dependence on the drying time peiord; see Fig.~\ref{fig:moisture}.

\bibliographystyle{apsrev4-1}
\bibliography{Akiba_PRE2017_inpress.bib}

\end{document}